\documentclass[preprint, superscriptaddress, nofootinbib, floatfix, 11pt]{revtex4-1} 
\pdfoutput = 1
\usepackage[colorlinks=true, linkcolor=blue, citecolor=blue]{hyperref}
\usepackage{graphicx}
\usepackage{graphics}
\usepackage{dcolumn}
\usepackage{bm}		
\usepackage{tabularx}
\usepackage{amsmath}
\usepackage{amssymb}
\usepackage{enumitem} 
\usepackage{verbatim}
\usepackage{hyperref}
\usepackage{cleveref}
\usepackage{lipsum}
\usepackage[normalem]{ulem}
\usepackage{braket}
\usepackage{physics}
\usepackage{color}
\usepackage{float}
\usepackage{natbib}
\usepackage{color}
\usepackage[dvipsnames]{xcolor}
\usepackage{xcolor}
\usepackage{xspace}
\usepackage{multirow}
\usepackage{appendix}
\usepackage[normalem]{ulem}	
\usepackage[utf8]{inputenc}
\newcommand{\beq}{\begin{equation}}
\newcommand{\baln}{\begin{aligned}}
\newcommand{\eeq}{\end{equation}}
\newcommand{\ealn}{\end{aligned}}
\newcommand{\beqn}{\begin{eqnarray}}
\newcommand{\eeqn}{\end{eqnarray}}

\newcommand{\pd}{\partial}
\newcommand{\thermal}{{\rm th}}
\newcommand{\plasma}{{\rm p}}
\newcommand{\Larmor}{{\rm L}}
\newcommand{\sh}{{\rm sh}}
\newcommand{\GeV}{{\rm GeV}}
\newcommand{\Debye}{{\rm D}}
\newcommand{\kms}{{\rm km/s}}
\newcommand{\Alfven}{Alfv\'{e}n }
\def\unitvec#1{{\hat{\bf #1}}}

\begin{document}
\count\footins = 1000
\interfootnotelinepenalty = 10000

\title{\Large Dynamics of millicharged dark matter in supernova remnants}
\author{Jung-Tsung Li}
\email{jul171@ucsd.edu}
\affiliation{Department of Physics, University of California, San Diego, California 92093, USA}
\author{Tongyan Lin}
\email{tongyan@physics.ucsd.edu}
\affiliation{Department of Physics, University of California, San Diego, California 92093, USA}
\date{\today}

\begin{abstract}
Dark matter could have a small electromagnetic charge, provided the charge-to-mass ratio is much less than that of electrons or protons. This candidate, commonly known as millicharged dark matter (mDM), would form a plasma and interact with the interstellar medium and electromagnetic fields within galaxies. In general, understanding the dynamics of mDM requires consideration of collective plasma effects. It has been proposed that mDM can be accelerated in supernova remnants, forming a dark cosmic ray population that would leave distinct experimental signatures. In this work, we study a microphysical model where mDM is shocked by a supernova remnant and isotropized in the frame of the expanding fluid. We find that for $|q_\chi/m_\chi| \gtrsim 10^{-13} e / {\rm MeV}$, the isotropization length for electromagnetic plasma instabilities is much shorter than the size of the supernova remnant. This is a necessary, though not sufficient, first step for formation of a Fermi-accelerated mDM component, and determining the size of this component requires further study. We discuss additional implications of mDM interactions in supernova remnants.
\end{abstract}

\maketitle

\makeatletter
\def\l@subsubsection#1#2{}
\makeatother
\tableofcontents

\newpage
\section{Introduction}

An important open question is whether dark matter (DM) has nongravitational interactions with Standard Model (SM) particles. In traditionally favored classes of DM candidates such as the WIMP, axion, or sterile neutrino, the DM typically has short-range interactions with the SM through mediators at the weak scale (or heavier). Such candidates and interactions are being actively searched for in indirect detection, direct detection and collider experiments.

If DM has long-range electromagnetic (EM) interactions with the SM, there can be dramatic effects on astrophysical and cosmological scales. Consider particle DM that has mass $m_\chi$ and a small electromagnetic charge $|q_\chi|$, with equal parts $+q_\chi$ and $-q_\chi$ components. We will refer to this candidate by the often-used nomenclature of millicharged DM (mDM), although possible values of $|q_\chi|$ range many orders of magnitude.\footnote{A DM candidate which is much heavier than the proton and where $|q_\chi|/e$ is an ${\mathcal{O}}(1)$ number is more often referred to as a charged massive particle (CHAMP)~\cite{DeRujula:1989fe,Dimopoulos:1989hk}.} For sufficiently small charge-to-mass ratios interactions, this DM candidate could have been undetected thus far. For instance, there are strong bounds from considering the scattering of mDM with the ionized plasma in the early universe, which leads to damping in the CMB anisotropies and matter power spectrum~\cite{Dubovsky:2003yn,McDermott:2010pa,Dvorkin:2013cea}. Current bounds~\cite{Kovetz:2018zan,Xu:2018efh,Slatyer:2018aqg} require that $|q_\chi|/e \lesssim 10^{-6} (m_\chi/\GeV)^{1/2}$, assuming DM mass $m_\chi$ below a GeV and that this candidate forms 100\% of the observed relic abundance.

Given such stringent constraints, what motivates our interest on the possibility of DM with fractional charge?  First, it provides a simple model where the DM relic abundance can be obtained through only EM interactions. For charges $|q_\chi/e| \sim 10^{-11}-10^{-10}$ it is possible to obtain the observed relic abundance through freeze-in~\cite{Chu:2011be,Dvorkin:2019zdi}, where the DM is never in thermal equilibrium with the SM thermal bath. (Note the constraints on $|q_\chi|$ exclude the possibility that 100\% of the DM came from thermal freeze-out of mDM.) This small fractional electric charge  could be generated if DM has a small fractional hypercharge, or if the DM is charged under a nearly massless dark photon which has a kinetic mixing with the SM photon (see for example Ref.~\cite{Davidson:1991si}).

The key idea is that even with such tiny charges, long-range EM interactions can give rise to observable signatures in experiments or in astrophysical environments. For instance, mDM has sizeable scattering rates in direct detection experiments, particularly through DM-electron scattering~\cite{Essig:2011nj}. Because the long-range interactions are enhanced for low momentum transfer scattering, a low-threshold experiment could probe values of $|q_\chi|$ where mDM is produced by freeze-in (or even smaller $|q_\chi|$). Direct detection experiments are now employing newly developed methods and technologies to search for this candidate and closely related models~\cite{Battaglieri:2017aum}.

In addition, mDM has been of interest following the reported observation of an anomalous absorption in radio frequencies by the EDGES collaboration~\cite{Bowman:2018yin}. The observation has been interpreted as a 21cm absorption trough due to DM-baryon scattering at cosmic dawn; since DM is typically much colder than baryons at these redshifts, this would cool the baryons relative to the CMB. Investigations have focused on a long-range Rutherford-type interaction such as that from mDM, since the scattering is larger at late times~\cite{Barkana:2018lgd,Barkana:2018cct,Berlin:2018sjs,Munoz:2018pzp}. For mDM comprising $100\%$ of the relic abundance, CMB bounds exclude the required charges to match the EDGES observation; however, there remain viable scenarios where mDM is only a fraction of the total DM~\cite{Kovetz:2018zan,Liu:2019knx}.

While the signatures discussed thus far focus on particle-particle interactions of mDM, the implications may be even more striking if we account for wave-particle interactions of mDM in galactic EM fields. Because of the tiny charges, mDM generally is not found in bound states and instead forms a dark plasma. In supernova remnants (SNR), this dark plasma can interact with the shock front and it has been suggested~\cite{Dimopoulos:1989hk,Chuzhoy:2008zy} that there is an mDM component that undergoes diffusive shock acceleration (DSA), also known as first-order Fermi acceleration~\cite{Bell:1978is, Blandford:1978bg}. Implicit in this scenario is the assumption that mDM efficiently scatters against the turbulent magnetic fields generated in a SNR. If this is the case, it was argued that the accelerated mDM is evacuated from the disk~\cite{Chuzhoy:2008zy} (which would hinder their detection on Earth) or alternatively that there is a flux of the accelerated component at Earth~\cite{Hu:2016xas,Dunsky:2018mqs} (which would aid in their detection). Mergers of mDM halos could result in the formation of collisionless shocks through plasma instabilities, which would impact galaxy formation as well as observations of cluster mergers such as the Bullet cluster; a related scenario where DM has long-range dark photon interactions was studied in Refs.~\cite{Heikinheimo:2015kra,Sepp:2016tfs}. Finally, it has also been pointed out that the interactions of halo mDM passing through the magnetic field of the Milky Way disk would lead to angular momentum exchange and result in spin-down of the disk~\cite{Stebbins:2019xjr}. Clearly, understanding such signatures and possible constraints is needed to determine the viability and detectability of mDM as a DM candidate.

Our goal is to study the consequences of the long-range electromagnetic interactions of mDM in the environment of supernova remnants. Specifically, in this work we will determine the mDM parameter space where mDM can be swept up by SNR and discuss possible implications. Previous work assumed that mDM can efficiently undergo DSA in remnants similar to proton cosmic rays (CR)~\cite{Chuzhoy:2008zy,Hu:2016xas,Dunsky:2018mqs}, but did not provide a microphysical justification or model. In fact, there are a number of stages for mDM dynamics in the SNR before they could be considered as undergoing DSA. The interaction of mDM with shocked ISM gas is a first necessary step, whereby the ambient mDM is swept up to $3/4$ of the shock speed and isotropized in the frame of the expanding fluid. Our approach is to develop a microphysical model for this process, where we can analyze the sweep-up timescale in the linear regime.

The basic idea is the following: viewed from the frame of the expanding SNR, the mDM plasma has a large bulk velocity and free energy, a configuration which is unstable to the generation of EM fields. If the growth times for these plasma instabilities are sufficiently fast, the bulk motion of the mDM is slowed down and it can become isotropized in the expanding remnant. Note that we will focus on the dynamics of the {\emph{bulk}} of the mDM. After the mDM is swept up by the SNR, it is possible that a small fraction of the mDM could cross the shock front multiple times and start the DSA process. However, obtaining a robust quantitative prediction for the fraction and spectrum of accelerated mDM requires additional techniques beyond the scope of this work.

A summary of the main results can be found in the remainder of this introduction. In Sec.~\ref{sec:proton_and_mDM}, we provide a review of supernova shock waves and acceleration of proton cosmic rays, make a comparison for mDM, and describe the plasma instabilities that we analyze. The detailed numerical results for the growth times are covered in Sec.~\ref{sec: ES instability} for electrostatic instabilities and in Sec.~\ref{sec: EM instability} for electromagnetic instabilities.  We then discuss some consequences for the mDM distribution in the Milky Way and for the evolution of SNRs in Sec.~\ref{sec:implications}, and conclude in Sec.~\ref{sec:conclusion}. Appendix~\ref{appendix: Landau dielectric function} briefly reviews the derivation of the linear response and growth rates in a plasma.

\begin{figure*}[t!]
\centering
    \includegraphics[width=0.73\textwidth]{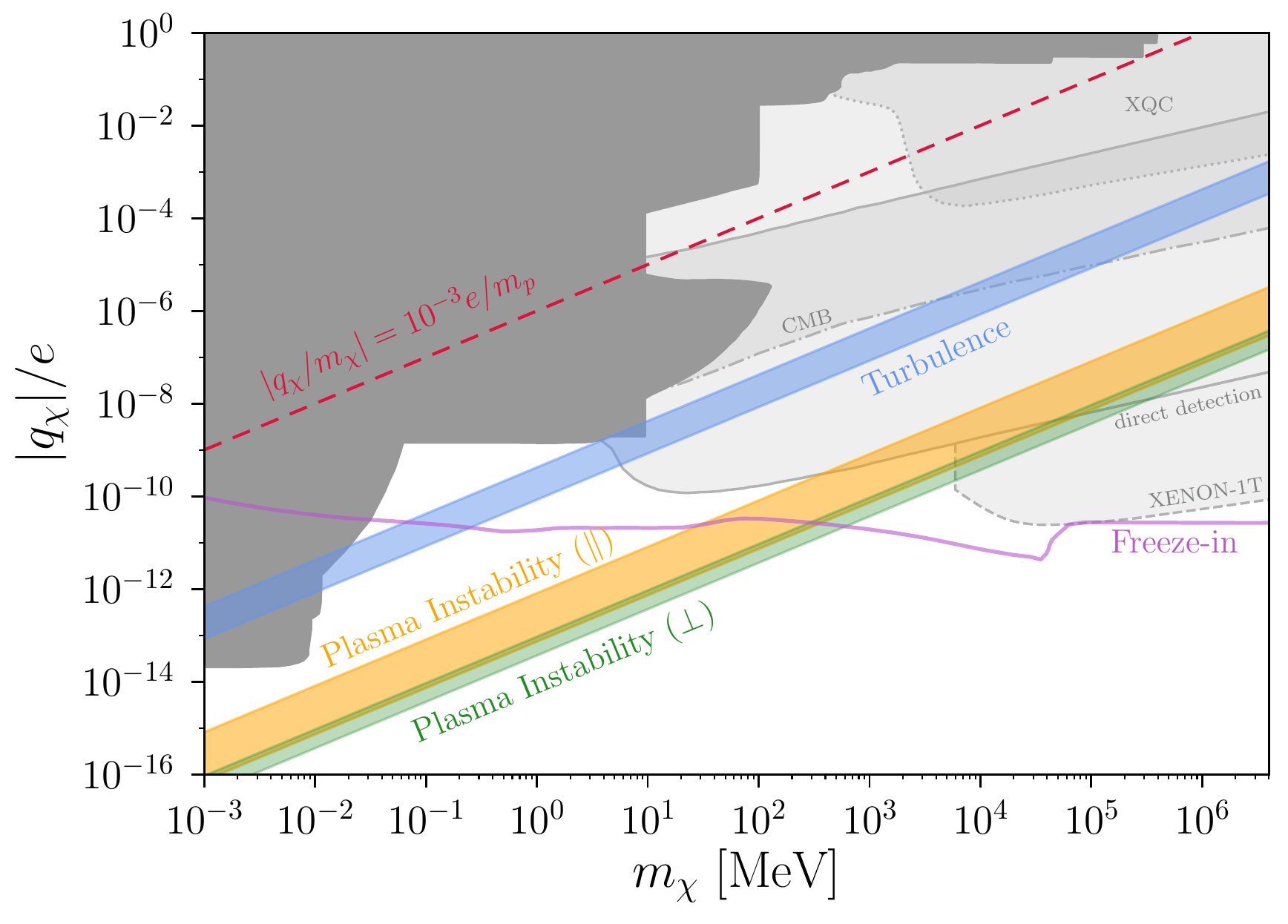}\\
    \caption{In this work, we only consider the parameter space below the red dashed line where $|q_\chi|/m_\chi < 10^{-3} (e/m_p)$. Above the colored bands, the mDM charge fraction $|q_\chi/e|$ is sufficiently large for it to be swept up by the SNR. The blue band gives the minimum $|q_\chi/e|$ for mDM diffusion due to CR proton-driven turbulence in a quasiparallel shock. Above the orange and green bands, there are mDM-driven plasma instabilities in parallel ($\parallel$) and perpendicular ($\perp$) shocks, respectively. The upper (lower) boundary of each band is for shock velocity $v_\sh = 1500~\kms$ ($300~\kms$). The dark gray shaded region combines bounds on production of mDM in accelerator experiments~\cite{Prinz:1998ua, Haas:2014dda, Magill:2018tbb, Jaeckel:2012yz}, in stars~\cite{Davidson:2000hf,Vogel:2013raa}, in SN1987a~\cite{Chang:2018rso}, and during BBN~\cite{Davidson:2000hf,Boehm:2013jpa, Foot:2014uba}. The lighter shaded regions are reported direct constraints on mDM that assume standard density and velocity distributions, which may be impacted by mDM wave-particle interactions. These include combined direct detection bounds~\cite{Emken:2019tni, Hambye:2018dpi}, from XQC~\cite{Mahdawi:2018euy}, and from CMB bounds on DM-baryon scattering~\cite{Xu:2018efh}.}
    \label{fig: mDM_parameter_space}
\end{figure*}

\subsection{Summary of results}

The main result of this work is illustrated in Fig.~\ref{fig: mDM_parameter_space}. We have assumed here that mDM is 100\% of the total DM relic abundance. We will only consider the parameter space below the red dashed line, where the mDM charge-to-mass ratio is at least $10^3$ times smaller than that of the proton, $|q_\chi|/m_\chi < 10^{-3} (e/m_p)$. Under this condition, the mDM Larmor frequency is at least $10^3$ times smaller than the proton Larmor frequency, so we can safely assume that mDM crosses the supernova shock front undeflected and that mDM does not change the dispersion relation of existing plasma waves in the Milky Way at a noticeable level. Our conclusions are:
\begin{enumerate}
\item For quasiparallel shocks (i.e., the angle between ambient magnetic field and shock normal is less than $45^\circ$) and $|q_\chi|/m_\chi$ above the blue band, it is possible for mDM to undergo pitch-angle scattering off CR-driven turbulent magnetic fields.  As discussed in Sec.~\ref{subsection: swept-up mDM by turbulence}, this is possible when the mDM Larmor radius does not exceed the maximum wavelength of CR-driven turbulence. A fraction of mDM particles might undergo Fermi acceleration by repeated scattering off the CR-driven turbulence upstream and downstream of the shock.
\item For quasiparallel shocks and $|q_\chi|/e$ between the blue and orange bands, there is a mDM-driven plasma instability. In this region, the mDM Larmor radius is sufficiently large that we approximate the magnetic fields as uniform. Then there is a plasma instability due to the large relative motion of the mDM and the expanding ionized fluid, which will act to reduce that relative velocity. In this part of the parameter space, the growth of the instability is sufficiently fast such that it saturates within one tenth of the SNR radius, and we treat the mDM as being effectively isotropized. A fraction of mDM particles could undergo Fermi acceleration by scattering off the mDM-driven turbulence.
\item For quasiperpendicular shocks, there is no Fermi acceleration of mDM. Similar to CR protons, the mDM cannot cross the shock multiple times in this case. However, above the green band, there is still a mDM-driven instability which affects the bulk dynamics of the mDM.
\end{enumerate}

While the strength of the plasma instability depends on the density of mDM and shock properties, roughly speaking the growth rate can be approximated by the Larmor frequency $\Omega_\chi$. The condition that the instability saturates within the SNR can therefore heuristically be given as  $v_{\sh}/\Omega_\chi \ll R_\sh$, where $R_\sh$ and $v_\sh$ are the shock radius and velocity, respectively. This can be rewritten as the condition that the mDM Larmor radius is much smaller than the shock size, $r_{{\rm L}, \chi} \ll R_\sh$, in agreement with na\"ive expectation. Our analysis demonstrates this condition robustly and takes into account the dependence of the instability on shock properties and ambient mDM density. In comparison to the wave-particle interactions, we note that everywhere below the red dashed line, i.e., $|q_\chi|/m_\chi < 10^{-3} (e/m_p)$, the mDM trajectory in the SNR would not be significantly altered by particle-particle Coulomb interactions. We can see this by writing the mean free path (mfp) of mDM-proton Coulomb scattering in the SNR~\cite{draine2010physics},
\beq
    {\rm mfp} = \frac{m_\chi^2 V_0^4}{8 \pi n_i q_\chi^2 e^2 \ln\Lambda} \approx 8100~{\rm pc} \, \left(\frac{10^{-3}}{\Omega_{\chi}/\Omega_{i}}\right)^2  \left(\frac{v_\sh}{300~\kms}\right)^4 \left(\frac{\rm 1~cm^{-3}}{n_i}\right) \left(\frac{25}{\ln \Lambda}\right),
    \label{eq: mfp coulomb}
\eeq
where $V_0 = 3v_\sh/4$ is the relative velocity between DM and shocked ISM, $n_i$ is the proton number density, and $\ln \Lambda$ is the Coulomb logarithm. This is much larger than the maximum SNR radius, which is $\lesssim {\rm 100~pc}$.

The darker gray shaded regions in Fig.~\ref{fig: mDM_parameter_space} show collected bounds on production of mDM, which do not make any assumption on its relic density. There are bounds on emission of mDM in stars, which would lead to anomalous cooling; we show limits from horizontal branch, white dwarf, and red giant stars for $m_\chi \lesssim 100$ keV~\cite{Davidson:2000hf,Vogel:2013raa} and from SN1987a for $m_\chi \lesssim 100$ MeV~\cite{Chang:2018rso}. For $m_\chi \lesssim$ MeV, mDM can thermalize with the SM thermal bath, leading to changes in BBN abundances and the effective number of light degrees of freedom; these bounds come from Refs.~\cite{Davidson:2000hf,Boehm:2013jpa}. Also included are collected accelerator bounds~\cite{Prinz:1998ua, Haas:2014dda, Magill:2018tbb, Jaeckel:2012yz}. The lighter shaded regions show constraints which make an assumption on the mDM density and/or velocity distribution. We show bounds assuming mDM is 100\% of the DM content from direct detection~\cite{Emken:2019tni}, searches for charged DM with the XQC satellite~\cite{Mahdawi:2018euy}, and the effect of DM-baryon scattering on the CMB~\cite{Xu:2018efh}. The solid line gives values of $|q_\chi|/e$  where 100\% of the DM relic abundance is comprised of mDM that is produced through freeze-in~\cite{Chu:2011be,Dvorkin:2019zdi}; values below this line lead to mDM that is a fractional component of DM, while larger $|q_\chi|/e$ requires additional interactions beyond EM.

Finally, in this paper we focus on a fermion mDM candidate with only EM interactions. As mentioned above, there are also models that generate the millicharge via a kinetically-mixed dark photon. DM interactions with the dark photon can lead to phenomenological differences in  constraints and DM interactions with charged SM particles. This possibility deserves study, but will not be considered any more here.

\section{Proton vs. millicharged dark matter dynamics in a supernova collisionless shock \label{sec:proton_and_mDM}}

In this section, we will overview the interaction between the supernova ejecta and the interstellar medium (ISM) and explain how mDM interacts differently with the supernova shock compared to the case of ISM gas. We will lay out the motivation and strategy for the plasma instabilities which we study in detail in Sec.~\ref{sec: ES instability} and~\ref{sec: EM instability}. Throughout this work, ``ion'' refers to ``proton'' unless otherwise specified.

\subsection{The visible sector: Protons and the supernova shock}
A supernova explosion is one of the most violent events in galaxies. A typical Type Ia supernova can expel $1~M_\odot$ ejecta with kinetic energy $\approx {\rm 10^{51}~erg}$ (i.e., with initial speed $\sim 10^4~{\rm km/s}$).\footnote{Type II core-collapse supernova can expel $10-20~M_\odot$ debris with kinetic energy as high as ${\rm 10^{52}~erg}$. For simplicity, we only consider Type Ia events in this work as a representative case.} Because the pressure in the ejecta is significantly higher than the pressure in the ambient interstellar medium, the ejecta will propel a shock wave to the ambient ISM. The ordinary matter bounded by this expanding shock wave--the ejecta from the explosion and the interstellar material swept up by the shock wave--is referred to as the \emph{supernova remnant}. As the interstellar gas is swept up by the shock, entropy is generated and the ordered bulk kinetic energy of the gas in front of the shock is converted to thermal energy. 

\begin{figure}[t!]
    \includegraphics[width=0.5\textwidth]{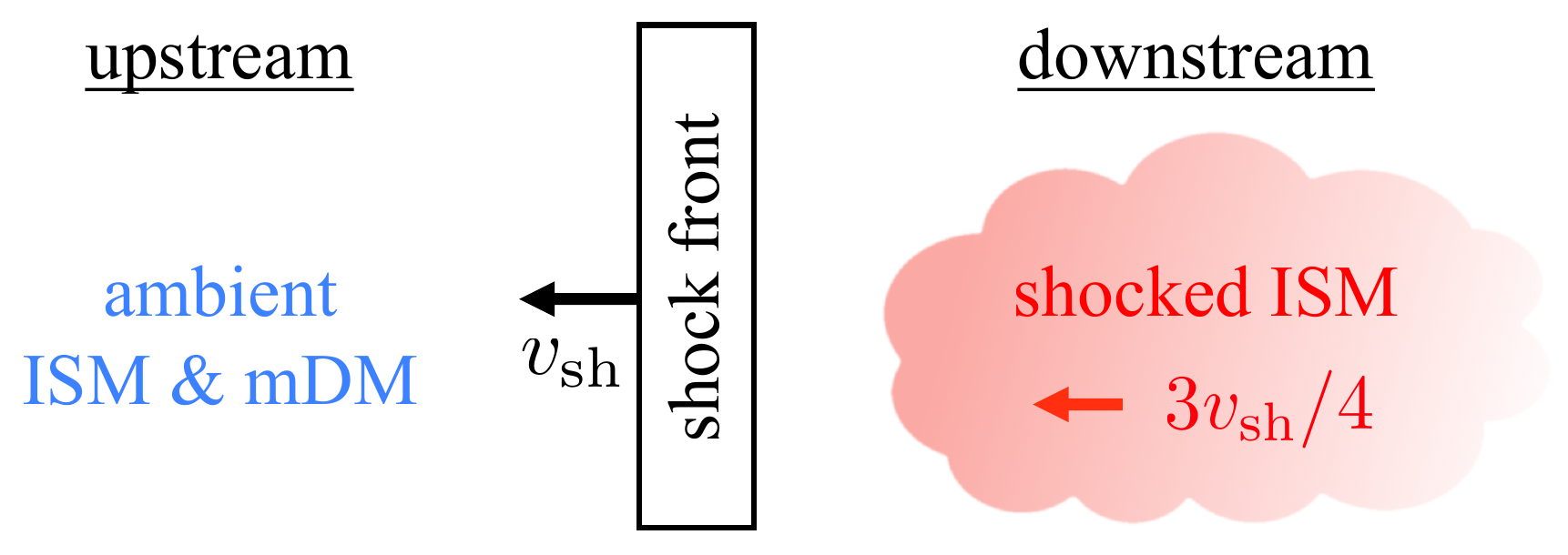}
    \caption{Schematic diagram for a supernova shock wave propagating in the interstellar medium (ISM). Just behind the shock front, the ISM has been shocked and moves at $3/4$ of the shock speed, whereas the mDM gas is at rest. The relative motion between the shocked ISM and the unshocked mDM provides the free energy to drive a plasma instability. If the instabilities occur, the mDM will be isotropized in the frame of the shocked ISM gas and be swept up by the expanding supernova remnant.}
    \label{fig: shock diagram}
\end{figure}

Figure~\ref{fig: shock diagram} is a schematic representation of a shock wave. The cooler, unshocked region ahead of the shock front is called \emph{upstream} and the hotter, shocked region is called \emph{downstream}. The shock transition zone is where the dissipation happens, and it requires the particle velocity distribution be isotropized in the downstream frame. We can determine the downstream fluid speed, density, and temperature from the Rankine-Hugoniot (RH) jump conditions, which relate the upstream and downstream states assuming the conservation of mass, momentum and energy in an one-dimensional flow. For a high Mach shock wave propagating in the monatomic gas, the downstream fluid speed in the frame of the background ISM is $3/4$ of the shock speed $v_\sh$. The number density and the magnetic fields of the shocked gas in the downstream are 4 times the number density and the magnetic fields in the upstream.

As for the downstream temperature, we take a somewhat more realistic case. Including helium, with a ratio of helium number density to hydrogen number density of 0.1, the downstream ion temperature is $T_{2i} \approx 3 \mu v_\sh^2 / 16$, where $\mu = 1.27 m_p$ denotes the mean mass per ion (Hydrogen and Helium nuclei) and $m_p$ is the proton mass. Thus, the proton thermal speed in the downstream is $v_{\thermal, i} \equiv \sqrt{{2 T_{2i}}/{m_p}} \approx 0.69 v_\sh$. We note that electrons have different temperature from ions just behind the shock front. This is because the ion-ion and electron-electron Coulomb scattering only leads ions and electrons to each relax to their own Maxwellian velocity distributions. The ion-electron relaxation time is much greater than ion-ion and electron-electron relaxation times, so the final temperature equilibration between ions and electrons happens at the deeper downstream.

Our discussion thus far requires the kinetic energy of the upstream ISM bulk flow to be converted into heat at the shock transition zone. For a supernova shock in interstellar space, it is collective plasma effects involving ions and plasma waves at the shock transition zone that cause the dissipation of the incoming ISM flow. This type of shock is referred to as a \emph{collisionless shock}~\cite{Draine:1993bc, burgess2015collisionless, Marcowith:2016vzl}. The formation and structure of the collisionless shock is complex as it involves several types of plasma instabilities and compression of the magnetic fields at the shock front. But in a broad brush, the collective plasma waves come from instabilities excited by a fraction of the ions reflected at the shock front. The thickness of the supernova collisionless shock is approximately a few ion Larmor radii, though it could be much larger if the ambient magnetic field is parallel to the shock normal~\cite{Draine:1993bc}.

In the following, we briefly review the evolution of supernova shocks and the generation of cosmic-ray protons. These will be important to understand the amount of mDM that can be affected by the shocks as well as the strength of preexisting turbulence in the downstream fluid.

\subsubsection{A brief history of a supernova remnant} 
In the early evolution of the supernova remnant, the shock wave propagates radially outward along with the supersonic ejecta at nearly constant speed, $\sim{\rm 10^4~km/s}$. This stage is called the \emph{free-expansion phase}. As the shock wave sweeps up more and more ambient gas, the swept-up mass eventually exceeds the ejecta mass and begins to govern the shock wave dynamics. Taking the number density of molecules in the ISM as ${\rm 0.25~cm^{-3}}$, this happens about $\sim330$~years after the explosion. 

\begin{figure}[t!]
\centering
    \includegraphics[width=0.48\textwidth]{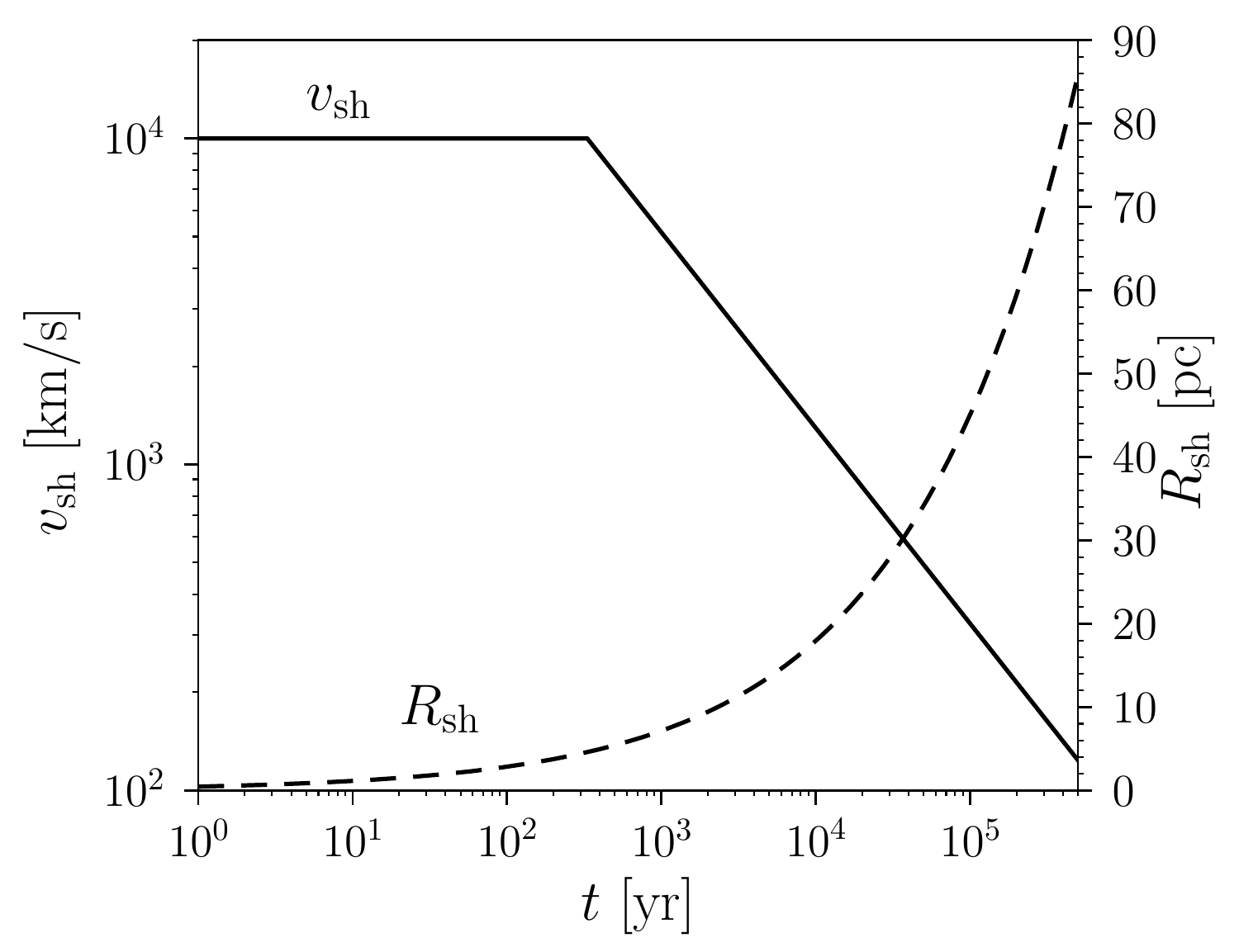}
    \caption{Typical expansion history for a Type Ia supernova remnant. The solid and dashed lines correspond to shock speed, $v_{\rm sh}$, and remnant radius, $R_{\rm sh}$, respectively. The free-expansion phase ends at $\approx 330$~years after the explosion.}
    \label{fig: remnant history}
\end{figure}

Subsequently, the shock starts to slow down, and the kinetic energy of the ejecta is transferred to the shocked matter. The shock now enters the \emph{Sedov-Taylor phase}. During this stage, the shock velocity and the distance it has traveled at time $t$ after the explosion are governed by the initial kinetic energy of the ejecta, $E_{\rm SN}$, and the mass density of the swept-up gas, $\rho_{\rm ISM}$. Without considering the radiative losses and ambient gas pressure, the variables $t$, $E_{\rm SN}$ and $\rho_{\rm ISM}$ should be the dominant quantities that control the dynamics of the shock. Using dimensional analysis, we can construct the time evolution of the shock radius and velocity as $R_\sh\left(t\right) = \kappa E_{\rm SN}^{1/5}\rho_{\rm ISM}^{-1/5}t^{2/5}$ and $v_\sh(t) = 0.4 \kappa E_{\rm SN}^{1/5}\rho_{\rm ISM}^{-1/5}t^{-3/5}$, respectively. For a monotonic gas, numerical calculations give $\kappa \approx 1.17$~\cite{shu1991physics}. In Fig.~\ref{fig: remnant history}, we show the history of a Type Ia supernova remnant. Eventually, at even later times, radiative losses become important and the SNR merges with the ISM.

\subsubsection{Diffusive shock acceleration of cosmic-ray protons} 
The standard picture for acceleration of cosmic-ray protons in the supernova shock relies on the presence of the interstellar magnetic field. This acceleration process is characterized by the angle $\theta$ between the shock normal and the background magnetic field. When $\theta$ is less (greater) than $45^\circ$, the shock is said to be quasiparallel (quasiperpendicular). For the special case of a shock with $\theta = 0^\circ$ ($90^\circ$), we call it a parallel (perpendicular) shock.

In a quasiparallel shock, a small fraction\footnote{Knowing the fraction of ions that are injected into the DSA process is one of the most difficult problems in cosmic ray physics. The standard acceleration theory utilizes the diffusion-convection equation~\cite{Skilling:1975id}. It requires isotropic pitch-angle distributions of the accelerated particles at the upstream and downstream of the shock and only works for particles with speed significantly greater than the shock speed~\cite{Malkov:1998dz}. The difficulty of determining the injection fraction is that the kinematics of low energy ions is extremely complicated as one has to simultaneously consider how the downstream thermal ions enter the upstream (or how the upstream ions are reflected by the potential barrier at the shock front~\cite{Caprioli:2014dwa}), drive a streaming instability, scatter at pitch angle, gain energy for the first few shock-crossings, escape the backstreaming \Alfven waves, and eventually diffuse across the shock front multiple times, all before being advected to the far downstream~\cite{Malkov:1995wd, Malkov:1998in}. This challenge is known as \emph{injection problem}.} of the charged particles can undergo the DSA process and be Fermi-accelerated to relativistic speeds. They scatter on the magnetic irregularities upstream and downstream and diffuse across the shock multiple times, gaining energy in each crossing. However, these magnetic irregularities are not preexisting, rather they are self-generated by the accelerated particles. The energetic cosmic rays streaming ahead of the shock have a highly anisotropic velocity distribution and therefore drive plasma instabilities~\cite{Bykov:2013lra} and excite \Alfven waves in the upstream of the shock. The waves drifting at the upstream move slower than the shock front and are later advected to the downstream and amplified by the shock. The downstream is expected to be highly turbulent. On the other hand, DSA is not operative in quasiperpendicular shocks because the charged particles do not propagate more than one thermal ion Larmor radius ahead of the shock front. As a result, there is very little magnetic turbulence generation, and the downstream magnetic fields are expected to be uniform in quasiperpendicular shocks. Such features have been seen in observations of SN 1006~\cite{Reynoso:2013tba} as well as in hybrid simulations of ion acceleration~\cite{Caprioli:2014tva, Hanusch:2019in}.

The characteristic lengths and amplitudes of the magnetic irregularities upstream and downstream can be inferred from the maximum energy of cosmic-ray protons. The observed cosmic-ray proton momentum spectrum is a power law with a nearly constant spectral index up to the ``knee'' energy, ${\rm 10^6~GeV}$, which indicates that cosmic-ray protons are Fermi-accelerated by one mechanism in each acceleration site--the DSA mechanism in supernova remnants~\cite{Zweibel:2013gw}. However, it was shown that considering the spatial dependence of the upstream diffusion coefficients and the finite lifetime of supernova remnants, the maximum energy of cosmic rays undergoing the DSA process is only ${\rm 10^4~GeV}$~\cite{Lagage:1983zz}. This result is obtained with the assumptions that (1) the upstream magnetic irregularities are driven by streaming instability and (2) the diffusion coefficient at the downstream is the Bohm-type, i.e., $\delta B/B \sim 1$. Since the turbulent magnetic fields are excited by cosmic rays, we expect the characteristic wavelengths of the excited waves should be comparable to the Larmor radius of the cosmic-ray protons. So the largest scale turbulence has a wavelength of approximately one Larmor radius of ${\rm 10^4~GeV}$ cosmic-ray protons. Assuming a downstream magnetic field of ${\rm 10~\mu G}$, the downstream  magnetic turbulence in quasiparallel shocks is present for length scales up to $\approx {\rm 10^{-3}~pc}$ with turbulence strength $\delta B/B \sim 1$.

To accelerate cosmic-ray protons to the knee energy in the supernova remnants, the diffusion coefficient has to be increased beyond the Bohm limit. This can be achieved if the turbulent magnetic fields are amplified to the level $\delta B/B \sim 10-100$. Such large magnetic turbulence likely is achieved by nonresonant hybrid instability in fairly young supernova remnants ($R\lesssim 1~{\rm pc}$)~\cite{Bell:2004cj}. However, younger supernovae cover less volume and so their impact in sweeping up mDM (which we will discuss in Sec.~\ref{subsection: swept-up mDM by turbulence}) is relatively small compared to the older remnants. For simplicity, we will only consider magnetic turbulence driven by cosmic-ray protons with energy up to $10^4$~GeV.

\subsection{The dark sector: mDM plunging to the downstream plasma}
The standard picture of the shock wave described above is restricted to the ions and electrons. We will take this as the leading dynamics for the supernova shock and treat mDM as test particles. Now we consider the dynamics of mDM particles as they enter the shock front. We will always work in the parameter space below the red dashed line in Fig.~\ref{fig: mDM_parameter_space} so that the mDM Larmor radius is always several orders of magnitude larger than the ion Larmor radius, $r_{\Larmor \chi}/r_{\Larmor i} \gg 1$.

Unlike the ion and electron flows which are dissipated at the shock transition zone through collective plasma effects, the mDM flow would not be isotropized in the same region. This is because the mDM Larmor radius is much larger than the width of the shock transition zone (about a few ion Larmor radii). Instead, the mDM should pass through the shock transition zone undeflected and plunge into the downstream ion-electron plasma. In the following, we will study how mDM particles interact with the downstream plasma so that they can be swept up by the supernova remnant. We will consider the possibility that mDM scatters on the magnetic turbulence driven by cosmic-ray protons as well as plasma waves driven by mDM particles.

\subsubsection{The diffusion of mDM in the quasiparallel shocks}\label{subsection: swept-up mDM by turbulence}

In the presence of magnetic irregularities driven by cosmic rays in the quasiparallel shock, mDM particles can undergo pitch-angle diffusion, which permits the mDM to be deflected by an $\mathcal{O}\left(1\right)$ angle and thus be swept up by the supernova remnant.
The diffusion happens when the mDM Larmor radius, $r_{\Larmor \chi} = m_\chi c V_{0}/|q_\chi| B$, is comparable to the wavelength of the magnetic irregularities, $\lambda$. Here $V_0 = 3v_\sh/4$ denotes the speed of the mDM flow in the rest frame of the downstream fluid. The associated mean free path for mDM pitch angle scattering\footnote{Note that the pitch angle diffusion and the sweep-up of mDM discussed here are restricted to the downstream of the quasiparallel shock. It is possible that the a small fraction of the charged particles can diffuse in the cosmic ray driven turbulent environment, cross the shock front multiple times, and start the DSA process. This idea has been applied to the dust grain acceleration in the supernova shocks~\cite{Epstein:1980dd, Ellison:1997kq}, and it could be a potential mechanism for cosmic-ray mDM acceleration.} through $90^\circ$ is~\cite{kulsrud2005plasma}
\beq
    \label{eq:mDM_turbulence}
    L_{\rm mfp} \approx \frac{r_{\Larmor\chi}} {\left(\delta B/B\right)^2}.
\eeq
To sweep up ambient mDM, (1) the pitch angle diffusion condition has to be satisfied, i.e., $r_{\Larmor\chi}\approx \lambda$, and (2) the mean free path cannot exceed the size of the supernova remnant.

As we discussed previously, turbulent magnetic fields driven by cosmic-ray protons in a  quasiparallel supernova shock exist for wavelengths up to $\lambda_{\rm max} \sim 10^{-3}$~pc with turbulence strength $\delta B/B\sim 1$. 
Thus, mDM with $r_{\Larmor \chi} < 10^{-3}$~pc can diffuse in the quasiparallel shock and $L_{\rm mfp}$ is approximately $r_{\Larmor\chi}$, in which case $L_{\rm mfp}$ does not exceed the size of the supernova remnant.
As a result, the condition that mDM can scatter with the magnetic turbulence and be swept-up by the quasiparallel shock is given as 
\beq
    \left(\frac{\rm 300~km/s}{v_\sh}\right)  \left(\frac{|q_\chi|/e}{8.4\times10^{-11}}\right)  \left(\frac{\rm MeV}{m_\chi}\right) > 1,
    \label{eq: swept-up by turbulence}
\eeq
where we assume the downstream magnetic field is $10~\mu{\rm G}$.

\subsubsection{Plasma instabilities in the supernova remnants}

As discussed above, mDM cannot undergo pitch angle scattering off cosmic-ray driven turbulence in some scenarios: (1) if the mDM Larmor radius is larger than $10^{-3}$~pc in a quasiparallel shock, or  (2) for any mDM Larmor radius in a quasiperpendicular shock, where there is little turbulence generated from cosmic-ray protons. For these cases, we will show instead that the mDM bulk flow can self-generate plasma waves and thus be swept-up by the expanding supernova remnants. Since in both cases the mDM Larmor radius is larger than any magnetic irregularities in the downstream plasma, we can approximately treat the  mDM as experiencing an ordered background magnetic field. The bulk of this work will then be devoted to analyzing possible mDM plasma instabilities in a homogeneous magnetic field.

In this section, we summarize the possible plasma instabilities that would allow a supernova shock to sweep up ambient mDM particles. In the frame of the shocked gas, we can treat the incoming mDM particles as a beam of charged particles moving with the bulk speed $V_0 = 3v_\sh/4$ and with the internal thermal speed approximately the Milky Way virial speed, $v_{\thermal,\chi} = v_{\rm vir} \approx 220~\kms$~\cite{Necib:2018igl}. The relative motion between the mDM beam and the shocked interstellar material then provides the free energy to drive plasma instabilities and excite plasma waves. Once the waves are excited, they will back-scatter on the mDM particles and slow down the mDM beam in the expanding fluid. That is, the mDM particles interact with the downstream fluid through wave-particle scattering. We assume the velocity distribution of the mDM particles becomes isotropized in the downstream frame when the instability saturates, and there is no more free energy to drive a plasma instability.

Unfortunately, there is no observational evidence guiding us as to which plasma waves and instabilities would be excited by mDM. For example, we do not know {\it a priori} the wave frequency, wavelength, and the propagation direction (parallel or perpendicular to $\mathbf{B}_0$). Nor do we know about the wave polarization (electrostatic vs. electromagnetic). In addition, there are several dozens of plasma waves and more than 50 kinds of plasma instabilities. As a starting point, here we will consider some of the most representative waves and instabilities that occur in the formation of astrophysical shocks~\cite{Ghosh:2000gc, Lesch:1990uh, Malkov:2018jy, Pohl:2000vh, Schlickeiser:2002dt, Nishikawa:2003us, Nishikawa:2004ug} as well as the anomalous heating in the pinch experiments~\cite{Krall:1971ey, Liewer:1973if, Davidson:1975kt, Gladd:1976by}.

\begin{table*}[ht]
\begin{ruledtabular}
\begin{tabular}{lcccll}
Instability   & Type & Beam direction                         & Wave direction                      & Frequency                      & Instability             \\ \hline
Ion-acoustic  & ES   & ${\mathbf V}_0 \parallel \mathbf{B}_0$ & $\mathbf{k} \parallel \mathbf{B}_0$ & $<\omega_{\plasma i}$                 & No (ion Landau damping) \\
Langmuir      & ES   & ${\mathbf V}_0 \parallel \mathbf{B}_0$ & $\mathbf{k} \parallel \mathbf{B}_0$ & $>\omega_{\plasma e}$                 & No ($V_0 < $ velocity threshold) \\
Lower-hybrid  & ES   & ${\mathbf V}_0 \perp \mathbf{B}_0$     & $\mathbf{k} \perp \mathbf{B}_0$     & $\sim\sqrt{|\Omega_i \Omega_e}|$ & No (ion Landau damping) \\ 
beam-firehose & EM   & ${\mathbf V}_0 \parallel \mathbf{B}_0$ & $\mathbf{k} \parallel \mathbf{B}_0$ & $ \lesssim |\Omega_\chi|$        & Yes                     \\ 
Weibel   & EM   & ${\mathbf V}_0 \perp \mathbf{B}_0$     & $\mathbf{k} \parallel \mathbf{B}_0$ & $ \lesssim |\Omega_\chi|$        & Yes                     \\ 
\end{tabular}
\end{ruledtabular}
\caption{The plasma instabilities studied in this work. Here $\mathbf{B}_0$ and $\mathbf{k}$ denote the ordered magnetic field in the downstream and the wave vector, respectively. The abbreviation ES and EM stand for electrostatic (longitudinal polarization) and electromagnetic (transverse polarization), respectively.}
\label{table: plasma instabilities}
\end{table*}

In Table~\ref{table: plasma instabilities}, we list the plasma waves and instabilities studied in this work. As a simplified model, we will take the downstream as a spatially homogeneous plasma immersed in an ordered magnetic field $\mathbf{B}_0 = B_0 \unitvec{z}$. We also assume the growth time of any mDM-driven instability in the supernova remnant is much greater than the dissipation time of interstellar protons and electrons at the shock front so that the mDM is treated as a beam drifting through the fully ionized hot proton/electron gas. With these assumptions, we find that the electrostatic waves are not excited because (1) the ion Landau damping dissipates the ion-acoustic and lower-hybrid waves, and (2) the mDM beam velocity is lower than the velocity threshold for exciting the Langmuir waves. On the other hand, electromagnetic waves may be excited in the low frequency regime ($\lesssim \Omega_\chi$). This is because the ion cyclotron frequency, $\Omega_i$, is much higher than the frequency of the mDM-driven electromagnetic waves, and thus ion cyclotron damping is avoided. Each of these instabilities is described in detail in Sec.~\ref{sec: ES instability} and~\ref{sec: EM instability}. 

We conclude this overview of mDM dynamics by providing a table of the notation used frequently throughout this work, Table~\ref{table: notation}. We will present equations in CGS-Gaussian units where the electric charge $e = \sqrt{\alpha \hbar c} = 4.8\times10^{-10}~{\rm statC}$. The relevant physical quantities for a particles species $j$ are the number density $n_j$, electric charge $q_j$, magnetic field strength in the shocked matter $B_0$, mass of the particle $m_j$, velocity of the particle $v_j$, temperature $T_j$, and mass density $\rho_j$. 

\begin{table*}[ht]
\begin{ruledtabular}
\begin{tabular}{llll} 
plasma frequency & $\omega_{\plasma j} \equiv  \sqrt{4\pi n_j q_j^2/m_j}$ & Larmor frequency & $\Omega_j \equiv q_j B_0/m_j c$ \\
Larmor radius & $r_{\Larmor j} \equiv  m_j c v_j/q_j B_0$ & Debye screening length & $\lambda_{\Debye j} \equiv \sqrt{T_j/4\pi n_j q_j^2}$ \\
thermal speed & $v_{\thermal,j} \equiv  \sqrt{2T_j/m_j}$ & \Alfven speed  & $v_{\rm A} \equiv \sum_j B_0/\sqrt{4\pi\rho_j}$ \\
\end{tabular}
\end{ruledtabular}
\caption{\label{table: notation} Notation used frequently in this work. We adopt CGS-Gaussian units and set $k_b = 1$.}
\end{table*}

\section{Electrostatic waves and instabilities}\label{sec: ES instability}
In this section, we study the possible electrostatic (longitudinal polarization) instabilities driven by an mDM beam. We will take a perturbative approach, where the mDM beam is treated as a perturbation to the remnant plasma, and determine the linear response. This approach is justified because the mDM plasma frequency is negligible compared to ion and electron plasma frequencies due to the small mDM electric charge. Given the parameter space we consider in this work, $|\Omega_\chi/\Omega_i| < 10^{-3}$, we are always working in the limit of small mDM plasma frequency, $\omega_{\plasma \chi}/\omega_{\plasma i} < 10^{-3} \sqrt{\rho_\chi/\rho_i} \ll 1$. With this assumption, the mDM does not change the dispersion relation of existing plasma waves in the Milky Way at a noticeable level.

While there are many kinds of electrostatic waves and instabilities, we choose three representative candidates as the most likely mDM instabilities: the ion-acoustic and Langmuir waves for $\mathbf{k}\parallel\mathbf{B}_0$ propagation and lower-hybrid waves for $\mathbf{k}\perp\mathbf{B}_0$ propagation. The reason we choose these three is motivated by their critical roles in the formation of astrophysical collisionless shocks (e.g., solar dust grain plasma~\cite{Ghosh:2000gc}, earth bow shock~\cite{Lesch:1990uh}, colliding stellar winds~\cite{Malkov:2018jy}) as well as the notorious turbulence heating in the theta-pinch experiments~\cite{Krall:1971ey, Liewer:1973if, Davidson:1975kt, Gladd:1976by}. However, while many astrophysical shocks have high beam velocity, the mDM beam velocity is similar to the downstream ion thermal velocity, which leads to an unavoidable strong Landau damping. As a result, we show below that these electrostatic waves are not excited by the mDM beam.

\subsection{Parallel shock (ion-acoustic and Langmuir waves)}
\begin{figure}[t]
    \centering
	\includegraphics[width=0.4\textwidth]{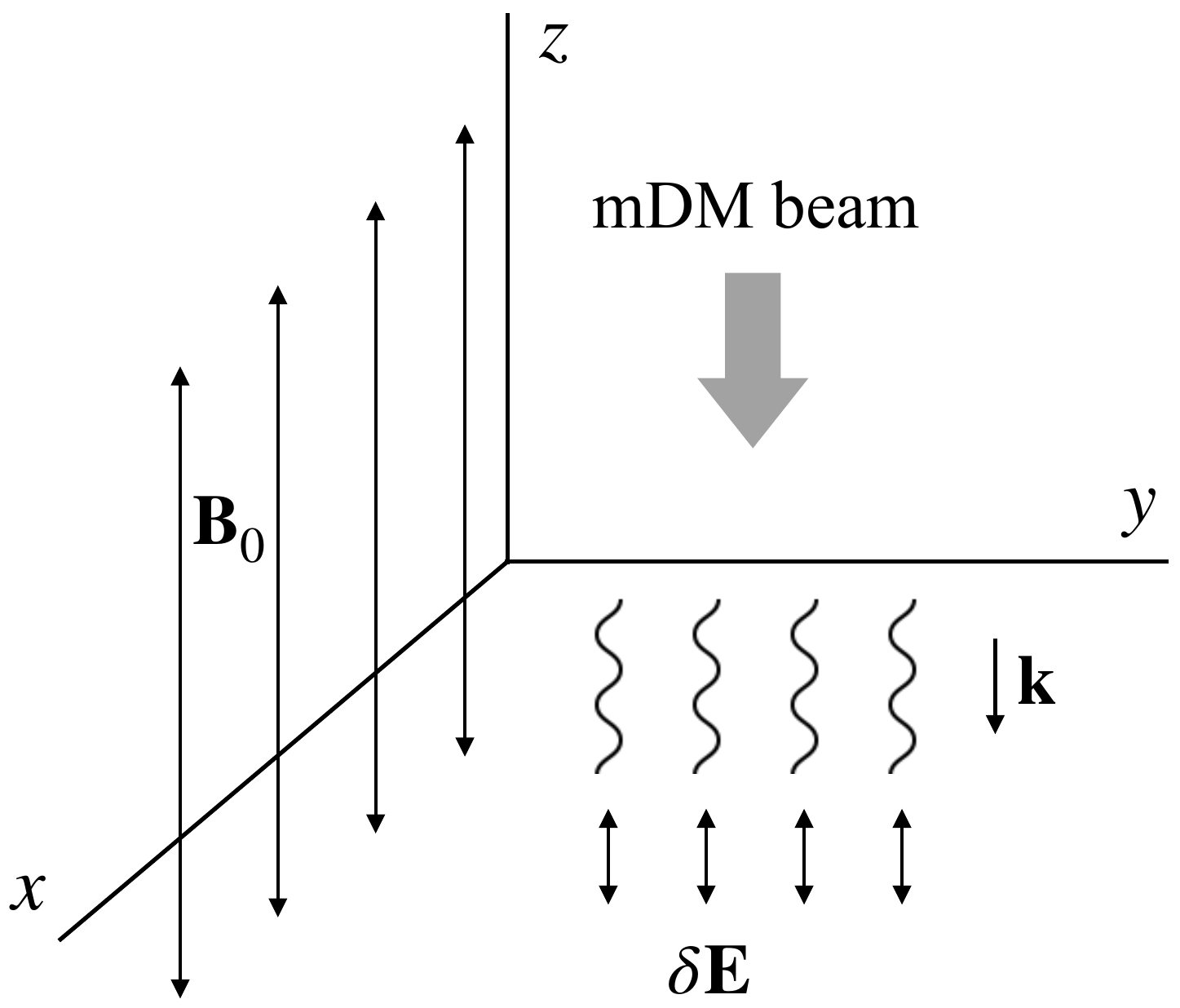}\\
	\caption{The environment in the downstream plasma frame, and our setup for a parallel shock where ${\mathbf B}_0 = B_0 \unitvec{z}$ is parallel to shock normal.  In this frame, the shock front propagates with velocity $(v_\sh/4)\unitvec{z}$ and the ambient mDM is treated as a weak beam flowing with velocity  $\mathbf{V}_0 \equiv -(3v_\sh/4)\unitvec{z}$. The ion-acoustic and Langmuir waves driven by the mDM beam propagate parallel to ${\mathbf B}_0$.} 
	\label{fig: diagram ES wave in parallel shock}
\end{figure}

We first consider the parallel shock scenario, defined as $\mathbf{B}_0 \parallel \mathbf{V}_0$, where $\mathbf{V}_0$ is the relative drift velocity between the mDM and the downstream electron-proton plasma. The setup is shown in Fig.~\ref{fig: diagram ES wave in parallel shock}. We examine the possibility of an mDM beam driving ion-acoustic and Langmuir wave instabilities. We choose these two representative electrostatic plasma waves for the following reason. In a parallel shock, the electrostatic waves are most easily excited when $\mathbf{k} \parallel \mathbf{V}_0$ and we therefore restrict to the case of parallel propagation, $\mathbf{k}\parallel \mathbf{B}_0$. Then the motions of the charged particles ($e^-$, $i^+$ and $\chi^\pm$) associated with this wave excitation/perturbation are parallel to $\mathbf{B}_0$, and the magnetic field does not alter the trajectories of these particles. As a result, electrostatic waves with $\mathbf{k} \parallel \mathbf{B}_0$ propagation would have dispersion relation identical to that of electrostatic waves in a (magnetic)field-free plasma~\cite{krall1973principles}. The two electrostatic waves for the field-free plasma are ion-acoustic waves in the low frequency regime ($\omega < \omega_{\plasma i}$) and  Langmuir waves in the high frequency regime ($\omega > \omega_{\plasma e}$), and so we study these waves.

Here we consider an mDM beam flowing through a background plasma consisting of Maxwellian electrons and protons. The normalized velocity distribution of each species is 
\begin{subequations}
\label{eq: distribution functions}
\begin{align}
    F_{0j} \left(\mathbf{v}\right) &=  \left(\frac{m_j}{2\pi T_j}\right)^{3/2} \exp\left(-\frac{m_j |\mathbf{v}|^2}{2T_j}\right), \quad j = e^-, i^+,
    \label{eq: distribution function, e i, thermal}\\
    F_{0\chi} \left(\mathbf{v}\right) &= \left(\frac{m_\chi}{2\pi T_\chi}\right)^{3/2} \exp\left(-\frac{m_\chi }{2T_\chi}\left(\mathbf{v}-\mathbf{V_0}\right)^2\right).
    \label{eq: distribution function, chi v, thermal}
\end{align}
\end{subequations}
We take the mDM thermal velocity to be the virial speed of Milky Way halo, i.e., $v_{\thermal,\chi} = \sqrt{2T_\chi/m_\chi} \approx v_{\rm vir} \approx 220~{\rm km/s}$. 

We take the standard approach for a linear stability analysis, wherein we determine the dielectric function in the presence of these species and study the imaginary part induced by the mDM beam.
In Appendix~\ref{subsec: plasma dispersion function}, we provide a detailed derivation of the dispersion relation for electrostatic waves (that is, longitudinal polarization with $\mathbf{k} \parallel \delta \mathbf{E}$) and parallel propagation, $\mathbf{k} \parallel \mathbf{B}_0$. Electrostatic waves have a dispersion relation determined by the poles of the dielectric function,
\beq
    0 = D\left(\omega_r + i \gamma, \mathbf{k} \right) = 1 + \frac{2 \omega_{\plasma i}^2}{k^2 v_{\thermal,i}^2} \big[1 + \xi_i Z\left(\xi_i\right)\big] + \frac{2 \omega_{\plasma e}^2}{k^2 v_{\thermal,e}^2} \big[1 + \xi_e Z\left(\xi_e\right)\big] + \frac{2 \omega_{\plasma \chi}^2}{k^2 v_{\thermal,\chi}^2} \big[1 + \xi_\chi Z\left(\xi_\chi\right)\big],
    \label{eq: dispersion relation parallel shock}
\eeq
where we have written the frequency in terms of real ($\omega_r)$ and imaginary ($\gamma$) parts, and $\xi_i$, $\xi_e$ and $\xi_\chi$ are defined by
\beq
    \xi_{i} = \frac{\omega_r + i \gamma}{k v_{\thermal, i}}, \quad \xi_{e} = \frac{\omega_r + i \gamma}{k v_{\thermal, e}}, \quad \xi_{\chi} = \frac{\omega_r - \mathbf{k}\cdot \mathbf{V}_0 + i \gamma}{k v_{\thermal, \chi}},
\eeq
with $\mathbf{k}\cdot \mathbf{V}_0 = k V_0$ and $v_{{\rm th},j} = \sqrt{2T_j/m_j}$. The function $Z\left(\xi_j\right)$ is referred to as the \emph{plasma dispersion function}, and defined explicitly in Eq.~(\ref{eq:Zdef}). The first three terms in Eq.~(\ref{eq: dispersion relation parallel shock}) support standard ion-acoustic and Langmuir waves. The last term, the dark matter contribution, can be neglected in determining the real oscillation frequency $\omega_r$ of the plasma waves  because we are working in the limit of $\omega_{\plasma \chi}/\omega_{\plasma i} \ll 1$. However, it is essential to include this term when calculating the growth rate $\gamma$ of the plasma waves, since the mDM beam is the only source of excess kinetic energy for driving an instability. Both the physics and approach here is reminiscent of the well-studied \emph{bump-on-tail instability} where an electron beam drifting with high velocity excites a Langmuir wave. In the following, we consider the possibility of having an instability in the ion-acoustic and Langmuir wave frequency regimes.

\subsubsection{Ion-acoustic waves}\label{subsubsec: ion-acoustic}
The ion-acoustic wave is a type of longitudinal oscillation in an unmagnetized plasma or in a magnetized plasma when $\mathbf{k} \parallel \mathbf{B}_0$. Its oscillation frequency is so low that the electrons are essentially locked to the oscillation of ions. The phase speed of the ion-acoustic wave is approximately ${\omega/k\sim\sqrt{(T_e + T_i)/m_i}}$, which is due to the restoring force of electron and ion thermal pressures. If ${T_e \sim T_i}$, the phase velocity is close to thermal ion velocity which suggests that a large fraction of ions can experience nearly constant electrostatic fields from the waves -- as if ions are ``surfing'' on them. Moreover, there is a large negative slope in the ion velocity distribution at the phase speed: physically, there are more thermal ions that are moving a bit slower than the phase speed, which takes away energy from the waves, compared to thermal ions moving slightly faster than the phase speed, which would give energy to the waves. The net effect is that the waves suffer from rapid energy loss due to ion Landau damping. On the other hand, if $T_e \gg T_i$ then the phase speed is on the tail of the ion distribution function, ${\omega/k \gg v_{\thermal,i}}$. Then there would be far fewer thermal ions ``surfing'' on the waves, and there is a reduced slope in the velocity distribution at the phase speed. The damping from ions is greatly suppressed in this case.

In a supernova shock environment, the electron and proton fluids are each isotropized after they cross the shock front. Each species has the same initial velocity in the downstream frame, which implies ${T_e / T_i \approx m_e/m_i \approx 1/1836}$ just behind the shock. The electrons are then heated to several tenths of $T_i$ through plasma instabilities~\cite{cargill1988mechanism, Ghavamian:2013aka}, but the final electron-ion temperature equilibration is caused by the electron-proton Coulomb scattering and the process takes about ${\sim 100~{\rm years}}$~\cite{draine2010physics}. However, the fact that $T_e \lesssim T_i$ suggests ion-acoustic waves suffer strong ion Landau damping. It is therefore not possible for mDM to excite any ion-acoustic waves in the supernova shock environment.

In the following discussion, we will demonstrate the problem of strong ion Landau damping by taking the case $T_e \gg T_i$. Although this does not correspond to the temperature ratio in the SNR, it  allows us to obtain an analytic result and see that Landau damping prevents an mDM instability even in a scenario where reduced damping is expected. In the $T_e \gg T_i$ limit, the phase velocity is in the range $v_{\thermal,i} \ll \omega_r/k \ll v_{\thermal, e}$ which corresponds $\xi_e \ll 1$ and $\xi_i \gg 1$.  Using the asymptotic expansions for $Z(\xi_j)$ given in Eqs.~(\ref{eq: Z, small xi}) and (\ref{eq: Z, large xi}), the real part of the dispersion relation in Eq.~(\ref{eq: dispersion relation parallel shock}) becomes
\beq
    0 = D_{\rm Re}\left(\omega_r, k\right) \approx  1 - \frac{\omega_{\plasma i}^2}{\omega_r^2} + \frac{1}{k^2 \lambda_{\Debye e}^2},
\eeq
where $\lambda_{\Debye e} \equiv \sqrt{T_e/4\pi n_e e^2}$ is the electron Debye length. (The mDM contribution to $D_r$ is neglected since we are working in the limit $\omega_{\plasma \chi}/\omega_{\plasma i} \ll 1$.)
The phase velocity is 
\beq
    \frac{\omega_r}{k} \approx \sqrt{\frac{T_e/m_i}{ 1 + k^2 \lambda_{\Debye e}^2} },
    \label{eq: phase speed}
\eeq
where $\sqrt{T_e/m_i} \equiv c_s$ is the ion sound speed. The real oscillation frequency $\omega_r$ ranges from $k c_s$ for $k^2\lambda_{\Debye e}^2\ll 1$ to $\omega_{\plasma i}$ for $k^2\lambda_{\Debye e}^2\gg 1$.\footnote{Note that the assumption of $\omega_r/k \gg v_{\thermal, i}$ breaks down when $k^2\lambda_{\Debye e}^2 \gtrsim T_e/T_i$, and the ion Landau damping becomes strong again. In the limit $k^2 \lambda_{\Debye e}^2 \gg T_e/T_i$ where the phase velocity $\omega_r/k \ll v_{\thermal,i}, v_{\thermal,e}$, there is no collective electrostatic plasma waves since the thermal ions and electrons can travel for significantly more than one wavelength, $k^{-1}$, within one period, $\omega_r^{-1}$. That is, any formation of electrostatic plasma waves will be ``washed out'' immediately. In the following discussion we always work in the limit $k^2 \lambda_{\Debye e}^2 < T_e/T_i$.}

Next, to evaluate the growth rate for a wave with $\mathbf{k}\parallel \mathbf{B}_0$, we take the limit $|\gamma/\omega_r| \ll 1$ and use Eqs.~(\ref{eq: growth rate equation}) and~(\ref{eq: D_REIM}). Then the damping rate from electrons plus ions, $\gamma_{e+i}$, is 
\beq
\baln
    \frac{\gamma_{e+i}}{\lvert\omega_r\rvert} 
    &\approx -\sqrt{\frac{\pi}{8}} \:  \left(\frac{\lvert\omega_r/k\rvert}{\omega_{\plasma i} \lambda_{\Debye e}}\right)^3 \bigg[ \sqrt{\frac{m_e}{m_i}} \exp\left(-\frac{\omega_r^2/k^2}{v_{\thermal,e}^2}\right)   + \left(\frac{T_e}{T_i}\right)^{3/2} \exp\left(-\frac{\omega_r^2/k^2}{v_{\thermal i}^2}\right)  \bigg] \\
    &\approx -\sqrt{\frac{\pi}{8}} \frac{1}{\left(1 + k^2 \lambda_{\Debye e}^2\right)^{3/2}}\bigg[ \sqrt{\frac{m_e}{m_i}} +  \left(\frac{T_e}{T_i}\right)^{3/2} \exp\left(-\frac{T_e}{2T_i\left(1 + k^2 \lambda_{\Debye e}^2\right)}\right) \bigg] ,
    \label{eq: ion and electron Ladau damping rate}
\ealn    
\eeq
and the growth rate from mDM, $\gamma_\chi$, is
\beq
    \frac{\gamma_\chi}{\lvert\omega_r\rvert} \approx \sqrt{\pi} \left(\frac{\omega_{\plasma \chi}}{\omega_{\plasma i}}\right)^2 \bigg\lvert \frac{\omega_r/k}{v_{\thermal,\chi}}\bigg\rvert^3 \left(\frac{V_0}{\omega_r/k} -1 \right)   \exp\left(-\frac{\left(\omega_r/k - V_0\right)^2}{v_{\thermal,\chi}^2}\right).
    \label{eq: mDM growing rate}
\eeq
The total growth rate is $\gamma = \gamma_{e+i} + \gamma_\chi$. From Eq.~(\ref{eq: mDM growing rate}), we find that the necessary condition to get $\gamma_\chi > 0$ is $V_0 > \omega_r/k$. Rewriting the beam speed as $V_0 = 3v_\sh/4 \approx 1.1 \, v_{\thermal,i}$, and using Eq.~(\ref{eq: phase speed}), we can simplify this condition  as $k^2 \lambda_{\Debye e}^2 > T_e/2.4 T_i -1$. For plasmas with $T_e/T_i < 2.4$, all $k$ modes have $\gamma_\chi > 0$. For plasmas with $T_e/T_i > 2.4$, only the modes with $k\lambda_{\Debye e} > \sqrt{T_e/2.4 T_i -1 }$ have $\gamma_\chi > 0$.

\begin{figure}[t!]
    \centering
    \includegraphics[width=0.48\textwidth]{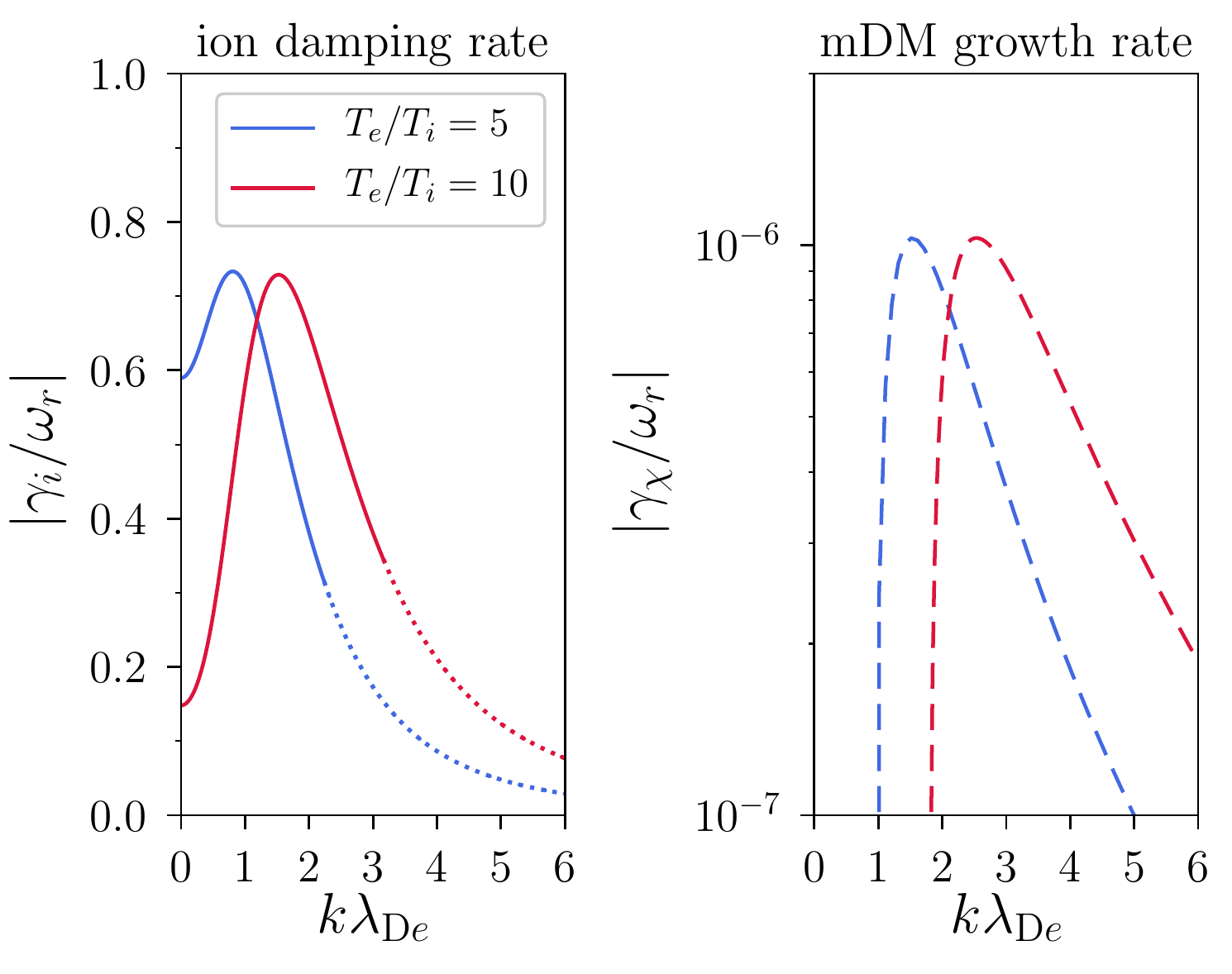}
    \caption{
    The damping of ion-acoustic waves from thermal ions (solid lines) from Eq.~(\ref{eq: ion and electron Ladau damping rate}), and growth from mDM (dashed lines) from Eq.~(\ref{eq: mDM growing rate}). Here we assumed $\omega_{\plasma \chi}/\omega_{\plasma i} = 10^{-3}$ and $v_\sh = 500~{\rm km/s}$. The dotted lines correspond to the regime $k^2\lambda_{\Debye e}^2 >  T_e/T_i$ where Eq.~(\ref{eq: ion and electron Ladau damping rate}) breaks down and there are no collective plasma oscillations. The damping rate from ions is significantly higher than the growth rate from mDM at all $k\lambda_{\Debye e}$. We have checked the same conclusion is true for shocks from the beginning of the supernova explosion ($v_\sh\sim 10^4~{\rm km/s}$) to the end of the Sedov-Taylor phase ($v_\sh\sim 200~{\rm km/s}$).
    } 
    \label{fig: ion_acoustic}
\end{figure}

Figure~\ref{fig: ion_acoustic} shows the damping rate from electrons plus ions, $\gamma_{e+i}$, and the growth rate from mDM, $\gamma_\chi$. We assume $\omega_{\plasma \chi}/\omega_{\plasma i} = 10^{-3}$ and $v_\sh = 500$ km/s, and consider plasmas with $T_e/T_i = 5$ and $10$. (Note that $\gamma_\chi$ is only plotted in the $k$ range where $\gamma_\chi > 0$.) It is clear from Fig.~\ref{fig: ion_acoustic} that damping dominates over  growth at all $k\lambda_{\Debye e}$, due in part to the strong suppression of the mDM contribution by $(\omega_{\plasma \chi}/\omega_{\plasma i})^2 < 10^{-6}$. We have checked the same conclusion is also true for shocks at the beginning of supernova explosion ($v_\sh\sim 10^4~{\rm km/s}$) to the end of the Sedov-Taylor phase ($v_\sh\sim 200~{\rm km/s}$). As a result, the mDM would not be swept up by the SNR via an ion-acoustic wave instability.

\subsubsection{Langmuir waves}
We now consider the Langmuir waves that exist in the large frequency regime ($\omega > \omega_{\plasma e}$). Langmuir waves, also known as electron plasma oscillations, are a type of fast-oscillating longitudinal wave in an unmagnetized plasma or in a magnetized plasma when $\mathbf{k} \parallel \mathbf{B}_0$. The oscillation is so rapid that the thermal ions are not able to catch up with the waves, and ions are essentially a static background, $m_i \rightarrow \infty$ and $|\xi_i| \rightarrow \infty$. As a result, the dispersion relation of the Langmuir waves is exclusively dictated by electrons. 

Similar to the discussion for ion acoustic waves, in order to avoid electron and ion Landau damping we require that the phase velocity $\omega_r/k \gg v_{\thermal, e},\: v_{\thermal, i}$. This corresponds to $\xi_e, \xi_i \ll 1$, and in this limit the real part of the dispersion relation in Eq.~(\ref{eq: dispersion relation parallel shock}) is given by
\beq
    0 = D_{\rm Re}\left(\omega_r, k\right) = 1 - \frac{\omega_{\plasma e}^2}{\omega_r^2} - 3 k^2 \lambda_{\Debye e}^2 \frac{\omega_{\plasma e}^4}{\omega_r^4} + \cdots,
    \label{eq: Langmuir dispersion relation}
\eeq
where again the mDM contribution is negligible in the limit $\omega_{\plasma \chi}/\omega_{\plasma i} \ll 1$. Solving Eq.~(\ref{eq: Langmuir dispersion relation}) gives the real wave frequency,
\beq
    \omega_r = \omega_{\plasma e} \sqrt{1 + 3 k^2 \lambda_{\Debye e}^2 + \cdots}.
    \label{eq: omega_r Langmuir waves}
\eeq
Note that these equations are only valid for $k \lambda_{\Debye e}  \ll 1$, which is required by the condition $\omega_r/k \gg v_{\thermal, e}$, and that Langmuir waves with $k \lambda_{\Debye e}  \sim 1$ suffer strong electron Landau damping.

The growth rate from mDM has the same form as Eq.~(\ref{eq: mDM growing rate}), but with the replacement $\omega_{\plasma i} \to \omega_{\plasma e}$ and with $\omega_r$ from Eq.~(\ref{eq: omega_r Langmuir waves}). And so again the condition for $\gamma_\chi > 0$ is $V_0 > \omega_r/k$. The phase speed in the regime $k \lambda_{\Debye e} \lesssim 1$ is approximately $\omega_r / k \gtrsim \omega_{\plasma e}\lambda_{\Debye e} = v_{\thermal,e}/\sqrt{2}$. As discussed earlier, the electrons are heated up to several tenths of $T_i$ behind the shock front due to plasma instabilities~\cite{cargill1988mechanism, Ghavamian:2013aka}, corresponding to $v_{\thermal,e} \sim  \sqrt{m_i/m_e} \, v_{\thermal,i} \sim 10\, v_{\sh}$. Then the phase speed is much larger than the mDM beam velocity in the downstream frame, $V_0$, and the condition for getting $\gamma_\chi>0$ is not satisfied. Therefore, the mDM would not be swept up by the SNR via a Langmuir wave instability.


\subsection{Perpendicular shock (lower-hybrid wave)}
\begin{figure}[t!]
    \centering
	\includegraphics[width=0.4\textwidth]{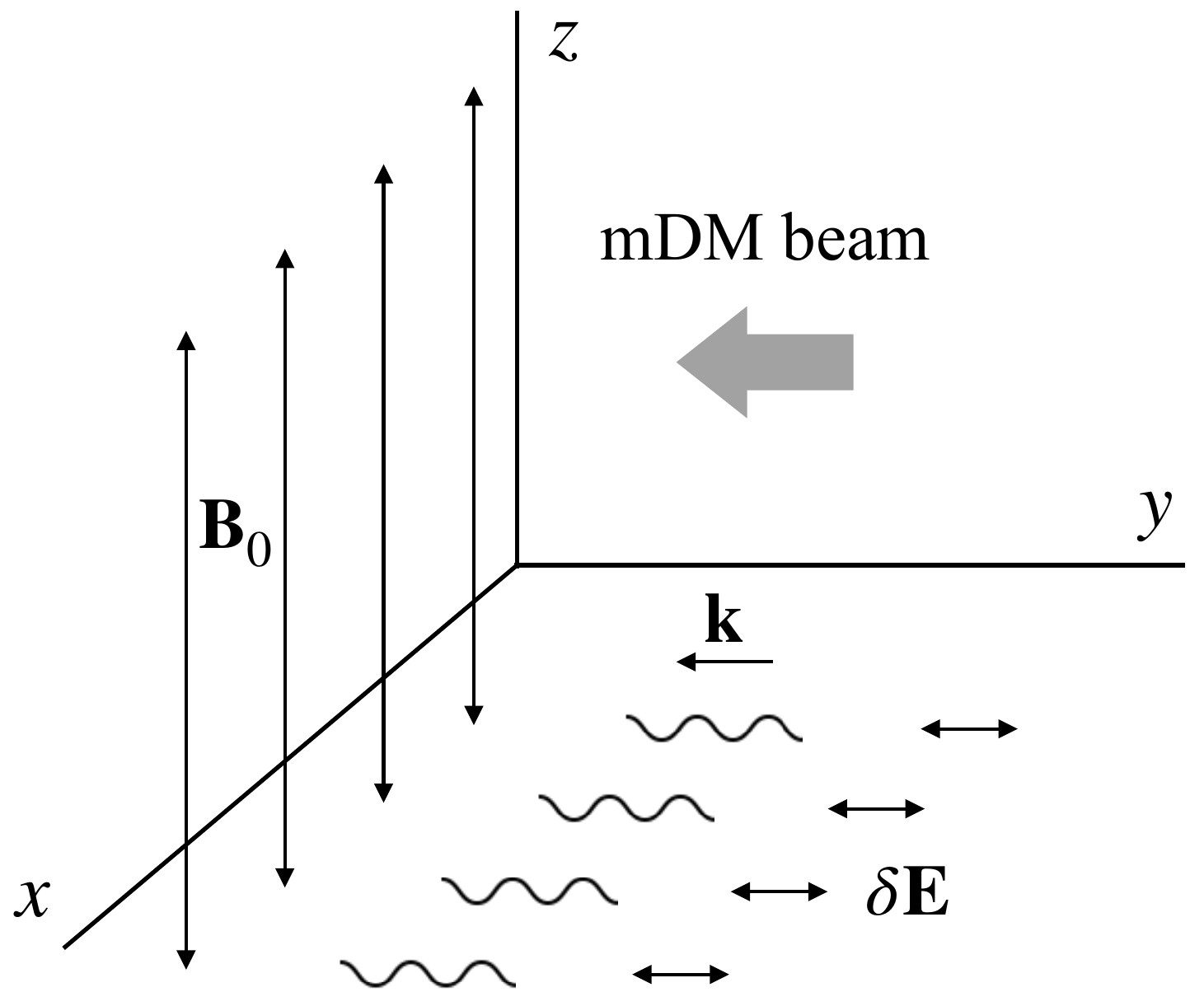}
	\caption{The environment in the downstream plasma frame, and our setup for a perpendicular shock where ${\mathbf B}_0$ is perpendicular to shock normal. In this frame, the shock front propagates with velocity $(v_\sh/4)\hat {\mathbf y}$ and the mDM is treated as a weak beam flowing with velocity ${\mathbf V}_0 \equiv -(3v_\sh/4)\hat {\mathbf y}$. The lower-hybrid waves driven by the mDM beam propagate perpendicular to ${\mathbf B}_0$. }
	\label{fig: diagram ES wave in perpendicular shock}
\end{figure}
For the perpendicular shock ($\mathbf{B}_0 \perp \mathbf{V}_0$), we choose lower-hybrid (LH) waves as the most likely candidate for an electrostatic wave driven by the mDM beam. The lower-hybrid wave instability is the high-frequency ($\omega \gg |\Omega_i|$) electrostatic wave driven by the cross-field plasma beam (i.e., moving perpendicular to the magnetic field $\mathbf{B}_0 = B_0 \unitvec{z}$) and the waves propagate very nearly perpendicular to $\mathbf{B}_0$. The free energy is provided either from the initial kinetic energy of the beam or from inhomogeneities in plasma density, temperature or background magnetic field. In this work, we only consider the first case since we do not expect the induced $\mathbf{E}$ and $\mathbf{E}\cross \mathbf{B}$ from local electron/proton charge separation to have significant effects on the mDM trajectories near the shock front. (The latter case, also known as \emph{lower-hybrid-drift instability}, is the major driver and free energy source for the anomalous heating in the theta pinch experiments as well as a dissipation mechanism in astrophysical shocks.)

\begin{figure}[t!]
    \centering
	\includegraphics[width=0.48\textwidth]{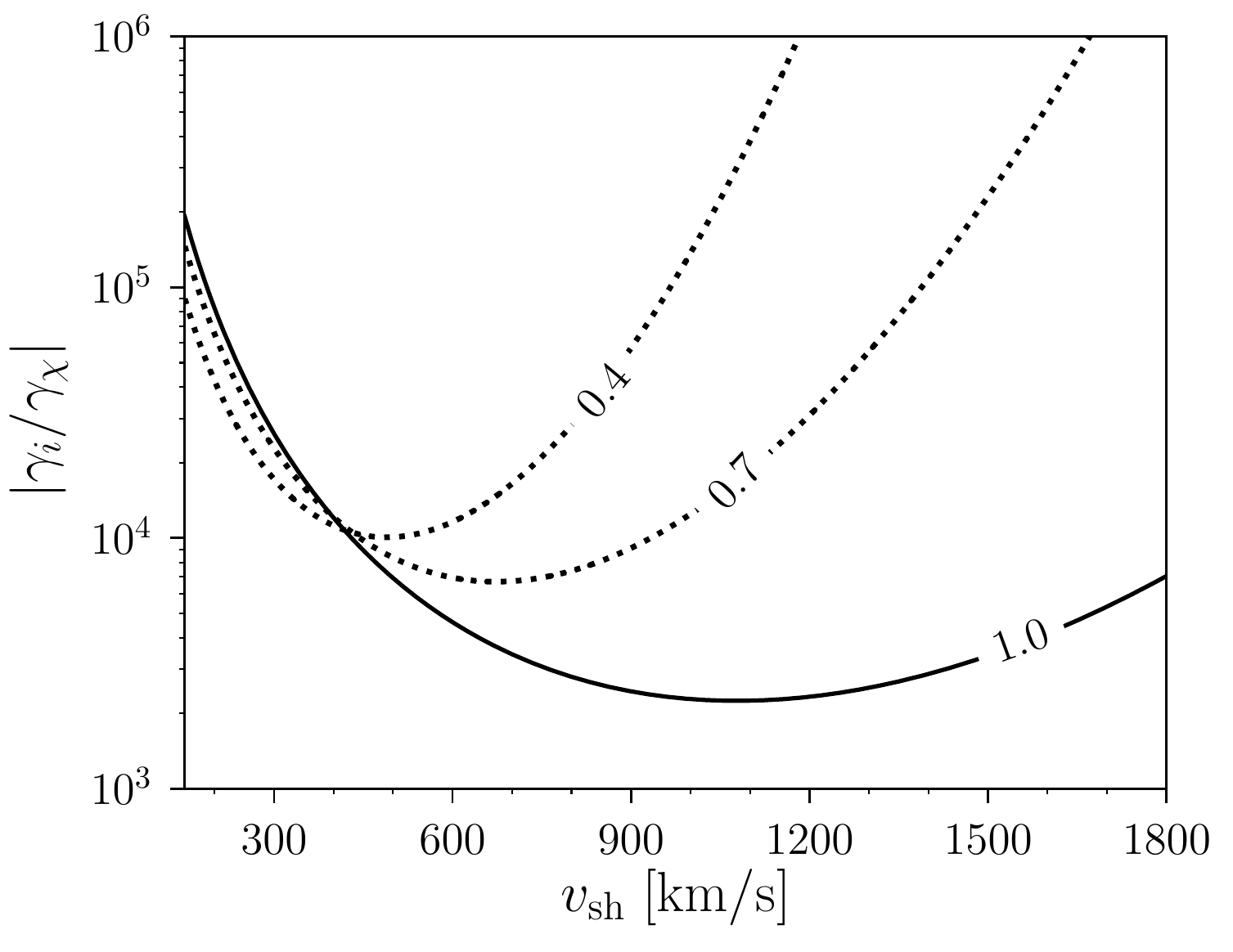}\\
	\caption{The ratio of ion damping to mDM growth rate for lower hybrid waves, $\lvert{\gamma_i}/{\gamma_\chi}\rvert$, as a function of shock speed. The various lines correspond to different fixed values of the ratio $\omega_r/kv_{\thermal,i}$. The curves are calculated under the assumption $\omega_{\plasma \chi} / \omega_{\plasma i} = 10^{-3}$. All the curves have $\lvert{\gamma_i}/{\gamma_\chi}\rvert \gg 1$, which indicates that the ion Landau damping will suppress any mDM beam-driven LH wave instability.}
	\label{fig: LH competition}
\end{figure}

The initial plasma configuration is illustrated in Fig.~\ref{fig: diagram ES wave in perpendicular shock}, and described as follows. We consider an mDM beam moving across the downstream magnetic field $\mathbf{B}_0 = B_0 \unitvec{z}$ with the beam velocity $\mathbf{V}_0 = V_0\unitvec{y}$ in the downstream frame. We focus on wave perturbations with $k_y^2 \gg k_x^2$ and $k_z=0$. Our choice is guided by the extensive parameter study of lower-hybrid-drift instability and modified-two-stream instability in Ref.~\cite{Gladd:1976by} where the author shows that the maximum instability growth of the LH waves occurs for $k_y^2 \gg k_x^2$, $k_z = 0$ and $k_y^2 r_{\Larmor e}^2 \approx 1$ when $T_e \approx T_i$. For simplicity, we write the wave vector as $\mathbf{k} = k \unitvec{y}$.

The typical frequency of the LH waves is around $\omega_r \sim \omega_{\rm LH} \equiv \omega_{\plasma i}/\sqrt{1 + \omega_{\plasma e}^2/\Omega_e^2}$. For ${\omega_{\plasma e}^2\gg \Omega_e^2}$, which is satisfied in the supernova downstream plasma, we have $\omega_{\rm LH} = \sqrt{m_i/m_e} |\Omega_i| \approx 43 |\Omega_i|$. The LH wave instability is characterized by magnetized electrons where the wave vector $k$ satisfies $k^2 r_{\Larmor e}^2 \approx 1$~\cite{Davidson:1975kt}. Since the thermal ion Larmor radius, $r_{\Larmor i} \approx 43 r_{\Larmor e}$, is larger than the characteristic wavelength of this instability, $1/k \approx r_{\Larmor e}$, we treat the ions as unmagnetized. The same argument applies to mDM. As a result, the dispersion relation is given as~\cite{Davidson:1975kt, Malkov:2018jy}
\beq
    0 = D\left(\mathbf{k},\omega \right) =  1 + \frac{2 \omega_{\plasma i}^2}{k^2 v_{\thermal, i}^2} \big[1 + \xi_i Z\left(\xi_i\right)\big] + \frac{\omega_{\plasma e}^2}{\Omega_e^2} \frac{1 - I_0\left(b\right)e^{-b}}{b} + \frac{2 \omega_{\plasma \chi}^2}{k^2 v_{\thermal, \chi}^2} \big[1 + \xi_\chi Z\left(\xi_\chi \right)\big],
    \label{eq: LH dispersion relation}
\eeq
where $b = k^2 r_{\Larmor e}^2/2 $, $\xi_i = \omega/k v_{\thermal, i}$, $\xi_\chi = \left(\omega - k V_0 \right)/k v_{\thermal, \chi}$ and $I_0\left(b\right)$ is the modified Bessel function of the first kind of order zero. Note that $I_0(b)$ is real and the magnetized electrons do not contribute to damping.

Because ions behave as an unmagnetized thermal gas within one LH wavelength, the ion Landau damping to the LH waves is non-negligible. Our goal is to determine whether the ion Landau damping will stabilize the plasma or if the mDM beam can successfully excite the LH waves. Utilizing Eqs.~(\ref{eq: growth rate equation}) and~(\ref{eq: D_IM}), the ratio of ion Landau damping to the mDM instability growth rate for LH waves is given as
\beq
\baln
    \frac{\gamma_i}{\gamma_\chi} &=  \frac{D_{\rm Im}^i\left(\mathbf{k},\omega_r\right)}{D_{\rm Im}^\chi\left(\mathbf{k},\omega_r\right)}\\ 
    &= - \left(\frac{\omega_{\plasma i}}{\omega_{\plasma \chi}}\right)^2 \left(\frac{v_{\thermal,\chi}}{v_{\thermal,i}}\right)^3 \left(\frac{\omega_r/k}{V_{0} - \omega_r/k}\right) \exp\left(-\frac{\omega_r^2/k^2}{v_{\thermal,i}^2} + \frac{\left(\omega_r/k - V_{0}\right)^2}{ v_{\thermal,\chi}^2}\right).
    \label{eq: LH, growth rate ratio}
\ealn
\eeq
Note that the necessary condition for $\gamma_\chi > 0$ is $\omega_r/k < V_0 = 3v_\sh/4 \approx 1.1 \, v_{\thermal, i}$, so we will always work in that condition. Taking the characteristic frequency of the LH wave, $\omega_r \sim \omega_{\rm LH} \approx  43 |\Omega_i|$, and the characteristic wave vector of the maximum growth rate of the LH wave instability, $k \sim 1/r_{\Larmor e}\approx 43 / r_{\Larmor i}$, the characteristic phase velocity is approximately $v_{\thermal, i}$. Of course, $\omega_r/k$ can be higher or lower than $v_{\thermal,i}$, as long as $\omega_r/k < 1.1 v_{\thermal,i}$. In Fig.~\ref{fig: LH competition}, we show $\lvert{\gamma_i}/{\gamma_\chi}\rvert$ as a function of shock speed at various values of $\omega_r/kv_{\thermal,i}$, assuming $\omega_{\plasma \chi} / \omega_{\plasma i} = 10^{-3}$. Note that decreasing $\omega_{\plasma \chi} / \omega_{\plasma i} $ would only raise the curves of $\lvert{\gamma_i}/{\gamma_\chi}\rvert$, meaning a weaker  growth rate from the mDM beam. Therefore, we find ion Landau damping is dominant over the growth rate from the mDM beam, and we do not expect mDM to be swept up by the supernova remnant via a LH wave instability.

\section{Electromagnetic waves and instabilities}\label{sec: EM instability}

In this section, we investigate the possibility of mDM exciting transverse electromagnetic (EM) waves in the downstream plasma, in the presence of a background magnetic field $\mathbf{B}_0$. We will show that an instability could occur with growth rate $\gamma \sim |\Omega_\chi|$. While the EM waves can, in general, propagate along arbitrary directions, we restrict our attention to wave propagation along $\mathbf{B}_0$ (${\bf k} \parallel \mathbf{B}_0$). Our choice is motivated by abundant observations of the enhanced magnetic fluctuations in the solar winds, which indicate that the dominant and the fastest growing EM instability modes propagate approximately parallel or antiparallel to $\mathbf{B}_0$~\cite{Kennel:1966jp, jian2014electromagnetic, Gary:2015em, Lazar:2017ex}. In that case, there are a variety of kinetic waves and associated instabilities, e.g., \Alfven waves with the firehose instabilities, ion cyclotron waves with the ion cyclotron instability, and electron cyclotron waves with the electron Whistler instability. Here we focus on transverse-polarization, parallel-propagating EM waves in a supernova shock. We then consider two limiting cases: a beam-firehose instability in a parallel shock (${\mathbf V}_0\parallel {\mathbf B}_0$) and a Weibel instability in a perpendicular shock (${\mathbf V}_0\perp {\mathbf B}_0$).

Similar to the approach of the previous section, here we solve for the linear dispersion relation for a transverse EM wave propagating along $\mathbf{B}_0 = B_0 \unitvec{z}$ with wave vector $\mathbf{k} = k \unitvec{z}$. Then the electric field perturbation can be written as $\delta \mathbf{E} = \delta E_x \unitvec{x} + \delta E_y \unitvec{y}$ and the general form of the linear dispersion relation can be expressed as~\cite{gary2005theory}
\beq
    0 = D^\pm \left(k, \omega \right) = c^2 k^2 - \omega^2 + \sum_j \omega_{\plasma j}^2 \int d^3 \mathbf{v} \frac{\left(\omega - k v_z\right) F_{0j} - \frac{1}{2} k v_\perp^2 \left({\partial F_{0j}  }/{\partial v_z}\right) }{\omega - kv_z \pm \Omega_j}.
    \label{eq: EM dielectric function general}
\eeq
The superscript of $D^\pm$ and the $\pm$ sign in front of $\Omega_j$ denote the right-handed ($+$, $\delta E_x = -i \delta E_y$) and left-handed ($-$, $\delta E_x = i \delta E_y$) EM waves. A cylindrical coordinate system for the velocity space is used and the integration is over the range $\int d^3{\mathbf v} = \int_0^\infty 2\pi v_\perp dv_\perp \int_{-\infty}^{\infty} dv_z $.


\subsection{Beam-firehose instability in a parallel shock}\label{Sec: firehose}

\begin{figure}[t!]
    \centering
	\includegraphics[width=0.4\textwidth]{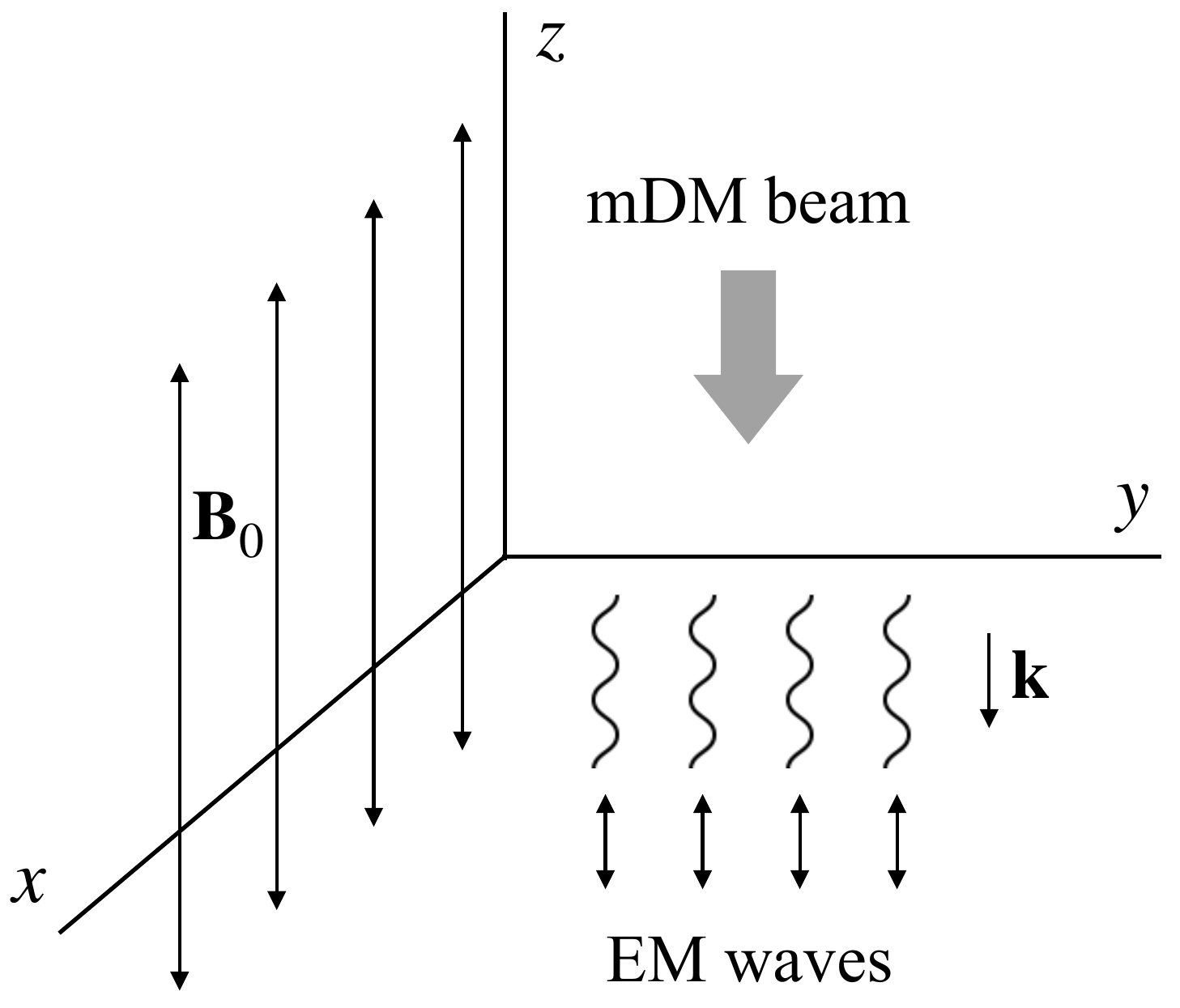}  \\
	\caption{The environment in the downstream plasma frame, and our setup for a parallel shock where ${\mathbf B}_0 = B_0 \unitvec{z}$ is parallel to shock normal. In this frame, the shock front propagates with velocity $(v_\sh/4)\unitvec{z}$ and the ambient mDM is treated as a weak beam flowing with velocity  $\mathbf{V}_0 \equiv -(3v_\sh/4)\unitvec{z}$. The excess of kinetic energy along $\mathbf{B}_0$ can drive a beam-firehose instability and excite EM waves.}
	\label{fig: diagram of beam-firehose}
\end{figure}

In the case of a parallel shock, we will show that the excess of kinetic energy from the mDM beam in the direction parallel to ${\mathbf B}_0$ can enhance EM perturbations and drive a beam-firehose instability.\footnote{Note that the instability studied here is slightly different from the classical firehose instability~\cite{osti_4329910}. In most of the literature, the firehose instability refers to the instability of the transverse EM waves propagating parallel or antiparallel to ${\mathbf B}_0$ in a two-temperature Maxwellian plasma system with $T_{j\parallel} > T_{j\perp}$ (where the subscripts $\parallel$ and $\perp$ refer to the directions relative to ${\mathbf B}_0$). The excited waves are in the \Alfven wave frequency regime, $|\omega|\ll |\Omega_i|$. On the other hand, the beam-firehose instability phrased in this work refers to a beam of mDM plasma flowing along ${\mathbf B}_0$ and driving an EM wave instability.} An illustration is shown in Fig.~\ref{fig: diagram of beam-firehose}. Physically, the firehose instability is caused by the back-reaction of the centrifugal force from charged particles moving along the curved magnetic lines ${\mathbf B} = \mathbf{B}_0 + \delta{\mathbf B}$. A small perturbation ${\delta \mathbf B}$ will cause charged particles to be redirected due to the Lorentz force, and a drift current is produced in the same direction as $\nabla \cross{\mathbf B}$. This drift current induces magnetic fields transverse to $\mathbf{B}_0$ and tends to increase the curvature of ${\mathbf B}$. The magnetic perturbations will enhance exponentially in time as the process is a coupled nonlinear feedback between the curved magnetic lines and the induced currents. This mechanism is reminiscent of a swinging firehose: when the water flow passes through the bent section of the hose, the centripetal force exerted on the hose tends to amplify the curvature and the bent section becomes even more curved. The process leads to an exponential growth of the wave amplitude~\cite{Parker:1958fg}. 

Now we consider the mDM beam flowing along ${\mathbf B}_0 = B_0 \unitvec{z}$ with a beam velocity in the downstream plasma frame written as ${\mathbf V}_0 = -(3v_\sh/4)\unitvec{z}$. The dispersion relation in this scenario is obtained by substituting the velocity distributions in Eqs.~(\ref{eq: distribution function, e i, thermal}) and~(\ref{eq: distribution function, chi v, thermal}) into the EM linear dispersion relation in Eq.~(\ref{eq: EM dielectric function general}), which gives
\beq
    0 = D^\pm \left(k, \omega \right) = c^2 k^2 - \omega^2 - \sum_{j = i^+, e^-} \omega_{\plasma j}^2 \left(\frac{\omega}{k v_{\thermal,j}}\right) Z\left(\xi_j\right) - \sum_{s = \chi^+, \chi^-} \omega_{\plasma s}^2 \left(\frac{\omega - k V_0}{kv_{\thermal,\chi}}\right) Z\left(\xi_{s}\right),
    \label{eq: firehose simplified dispersion relation}
\eeq
where $\xi_j = \left(\omega \pm \Omega_j \right)/kv_{\thermal,j}$ and $\xi_s = \left(\omega - k V_0 \pm \Omega_s\right)/ kv_{\thermal,\chi}$ and $Z(\xi)$ is again the plasma dispersion function defined in Appendix~\ref{subsec: plasma dispersion function}. Assuming equal number density of $\chi^+$ and $\chi^-$, we have $n_{0\chi^+}=n_{0\chi^-}=n_{0\chi}/2$ and consequently $\omega_{\plasma \chi^+}^2 = \omega_{\plasma \chi^-}^2 = \omega_{\plasma\chi}^2/2$. For $V_0 = 0$, there is no free energy to drive an instability and  Eq.~(\ref{eq: firehose simplified dispersion relation}) supports only stable solutions with $\gamma \leq 0$. Here we show that in the presence of the mDM beam with nonzero $V_0$, it is possible to excite unstable modes of a beam-firehose instability with characteristic wave frequency $|\omega| \lesssim \Omega_\chi$.

We first focus on the electrons and ions, which we assume to be strongly magnetized with $|\omega|\ll |\Omega_i|,|\Omega_e|$. In the long wavelength and low frequency limit that we are interested in, we have $|\xi_i|,|\xi_e| \gg 1$. Then we can approximate $Z\left(\xi_j\right) \approx -1/\xi_j$ for $j = i^+, e^-$ and simplify the ion and electron terms in Eq.~(\ref{eq: firehose simplified dispersion relation}):
\beq
   \sum_ {j = i^+, e^-}\frac{\omega_{\plasma j}^2 \omega}{\omega \pm \Omega_j} \approx \omega \bigg[\frac{\omega_{\plasma e}^2}{\pm \Omega_e} \left(1 - \frac{\omega}{\pm \Omega_e}\right) + \frac{\omega_{\plasma i}^2}{\pm \Omega_i} \left(1 - \frac{\omega}{\pm \Omega_i}\right) \bigg] = - \omega^2 \frac{c^2}{v_{\rm A}^2},
\eeq
where the charge neutrality of proton-electron plasma is used. The standard \Alfven speed $v_{\rm A}$ is defined as $v_{\rm A} \equiv B_0/\sqrt{4\pi\left(\rho_i + \rho_e \right)}$ and has a typical value
\beq
    v_{\rm A} \approx 22~{\rm km/s} \: \left(\frac{{ B}_0}{10~\mu{\rm G}}\right) \left(\frac{1~{\rm cm^{-3}}}{n_{{\rm post},i}}\right)^{1/2},
\eeq
where $n_{{\rm post},i}$ denotes the proton number density in the supernova downstream fluid.

\begin{figure}[t!]
\centering
    \includegraphics[width=0.49\textwidth]{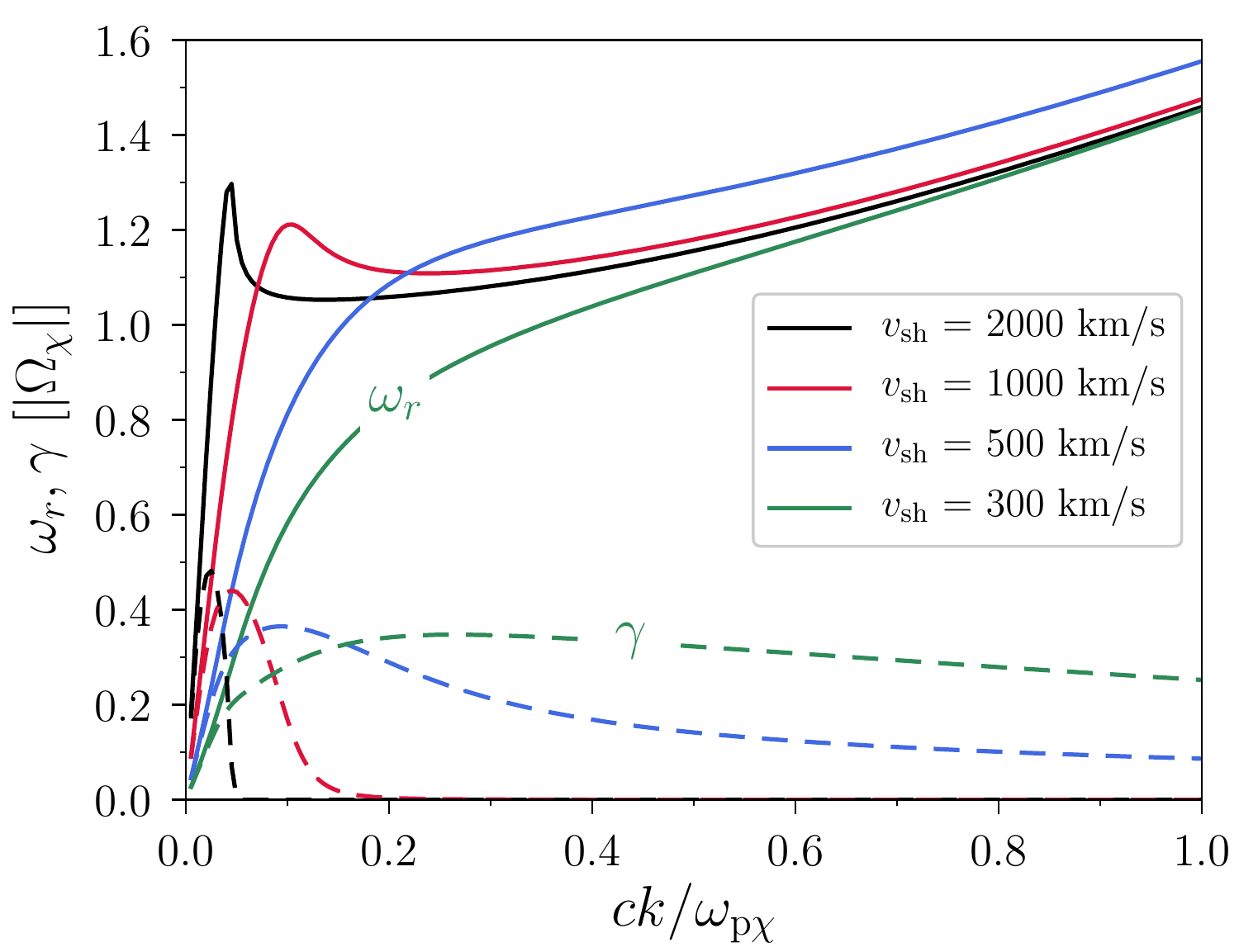}
    \includegraphics[width=0.49\textwidth]{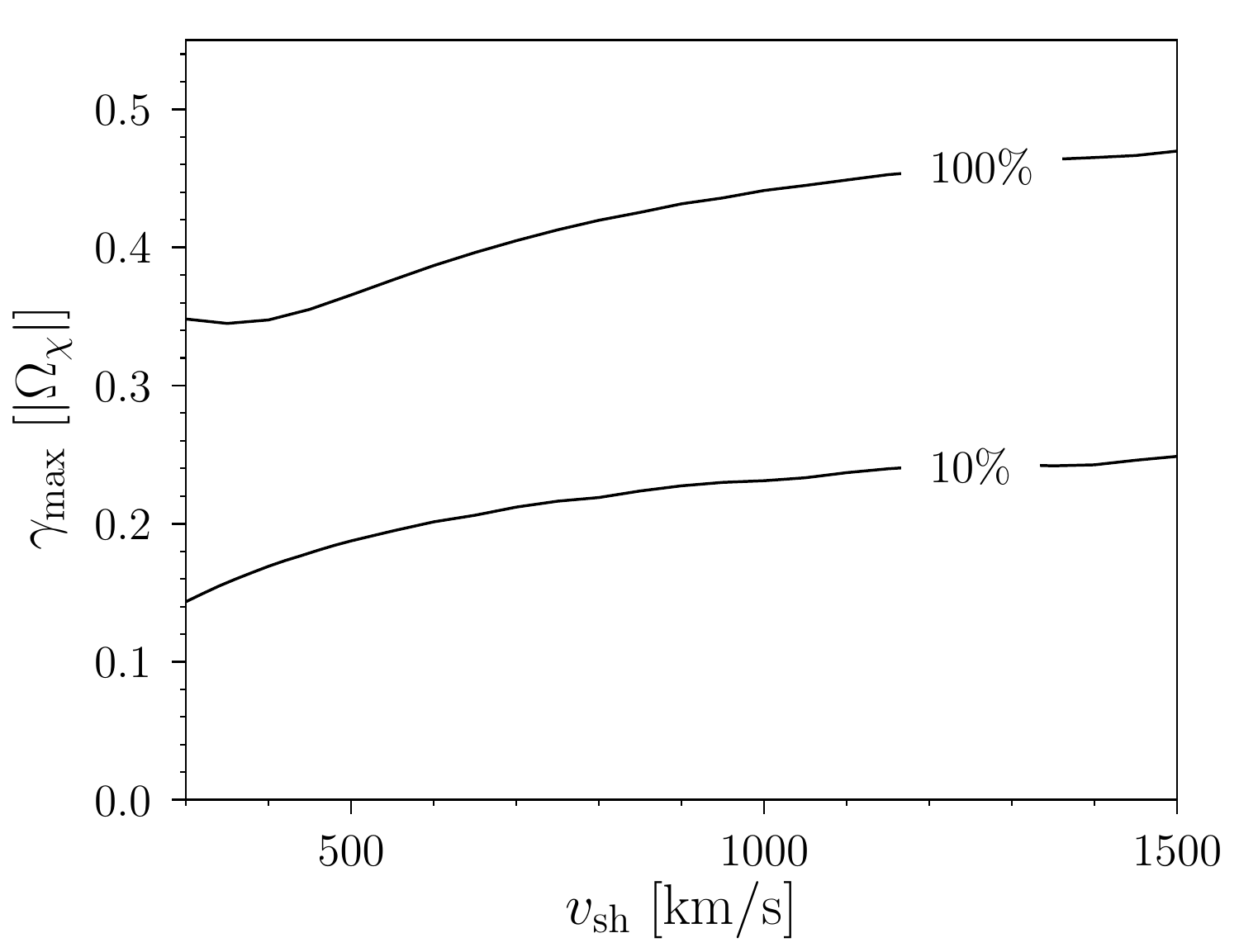}
    \caption{Beam-firehose instability in a parallel shock. (left) Plot of the real oscillation frequency (solid lines) and the growth rate (dashed lines) of the right/left-handed EM waves in units of $|\Omega_\chi|$. The curves are plotted assuming $\rho_{\rm DM} = \rho_{\chi} = {\rm 1~GeV/cm^3}$. (right) Plot of the maximum growth rate as a function of shock speed. The numbers on the curves denote the fraction of the DM abundance comprised of mDM,  $\rho_{\chi}/\rho_{\rm DM}$, assuming $\rho_{\rm DM} = {\rm 1~GeV/cm^3}$.}
    \label{fig: firehose_like_omega}
\end{figure}

The relevant source of damping for EM waves propagating in a homogeneous plasma is cyclotron damping. For the case studied here, the cyclotron damping from thermal ions and electrons is negligible. This is  because for very low frequency and long wavelength EM waves ($|\omega|\ll |\Omega_i|,|\Omega_e|$ and $|\xi_i|,|\xi_e| \gg 1$), the exponential function in the plasma dispersion function $Z\left(\xi_j\right)$ which contributes the damping is exponentially suppressed. Physically, it means the ions and electrons do not rotate at the same rate as the circularly polarized EM waves, so the waves do not undergo a cyclotron resonance. The ion (electron) cyclotron damping only becomes important when $\omega \to \Omega_{i}$ ($\Omega_{e}$), in which case the ions (electrons) can absorb a significant amount of the energy from the electric fields of the EM waves.

In the presence of the mDM beam, the dispersion relation of right-handed EM waves becomes
\beq
    0 = c^2 k^2 - \omega^2 \left(1 + \frac{c^2}{v_{\rm A}^2}\right) - \frac{\omega_{\plasma \chi}^2}{2} \left(\frac{\omega - k V_0}{k v_{\thermal,\chi}}\right) \bigg[ Z\left( \xi_{\chi^+} \right) + Z\left( \xi_{\chi^-} \right) \bigg].
    \label{eq: firehose final dispersion relation}
\eeq
The left-handed EM waves have an identical form since $\Omega_{\chi^+} = -\Omega_{\chi^-}$. We solve Eq.~(\ref{eq: firehose final dispersion relation}) numerically assuming $B_0 = 10~{\mu{\rm G}}$ and $n_{{\rm post},i} = 1~{\rm cm^{-3}}$.  The results are shown in Fig.~\ref{fig: firehose_like_omega}. In the left panel of Fig.~\ref{fig: firehose_like_omega}, we show the real oscillation frequency $|\omega_r|$ and the growth rate $\gamma$ as a function of $ck/\omega_{\plasma \chi}$ for different shock speeds, assuming $\rho_{\rm DM} = \rho_{\chi} = {\rm 1~GeV/cm^3}$. The value of $k$ at which the maximum growth occurs is a decreasing function of the shock speed $v_\sh = 4V_0/3$. This is because the instability is most effective when the crossing time of the beam through one wavelength, $\sim1/kV_0$, is comparable to one gyration time of $1/\Omega_\chi$, i.e., the beam is \emph{resonant} with the excited EM waves. This also justifies the assumption made above that $|\xi_i|, |\xi_e| \gg 1$ since $|\xi_i| = \left(\omega \pm \Omega_i\right)/k v_{\thermal,i} \sim |\Omega_\chi/kV_0||\Omega_i/\Omega_\chi| \sim |\Omega_i/\Omega_\chi| \gg 1$ and the same argument applies for the electrons as well. In the right panel of Fig.~\ref{fig: firehose_like_omega}, we plot the maximum growth rate $\gamma_{\rm max}$ as a function of $v_\sh$ for various values of $\rho_{\chi}/\rho_{\rm DM}$, assuming $\rho_{\rm DM} = {\rm 1~GeV/cm^3}$. The maximum growth rate $\gamma_{\rm max}$ is an increasing function of $v_\sh$ and $\rho_{\chi}/\rho_{\rm DM}$, but $\gamma_{\rm max}$ starts to saturate at high $v_\sh$. For the optimal scenario where all DM is made up of mDM, we find $\gamma \sim 0.5 |\Omega_{\chi}|$ for all shock speeds in the Sedov-Taylor phase ($v_\sh \gtrsim 200~\kms$).

\subsection{Weibel instability in a perpendicular shock}

\begin{figure}[t!]
    \centering
	\includegraphics[width=0.4\textwidth]{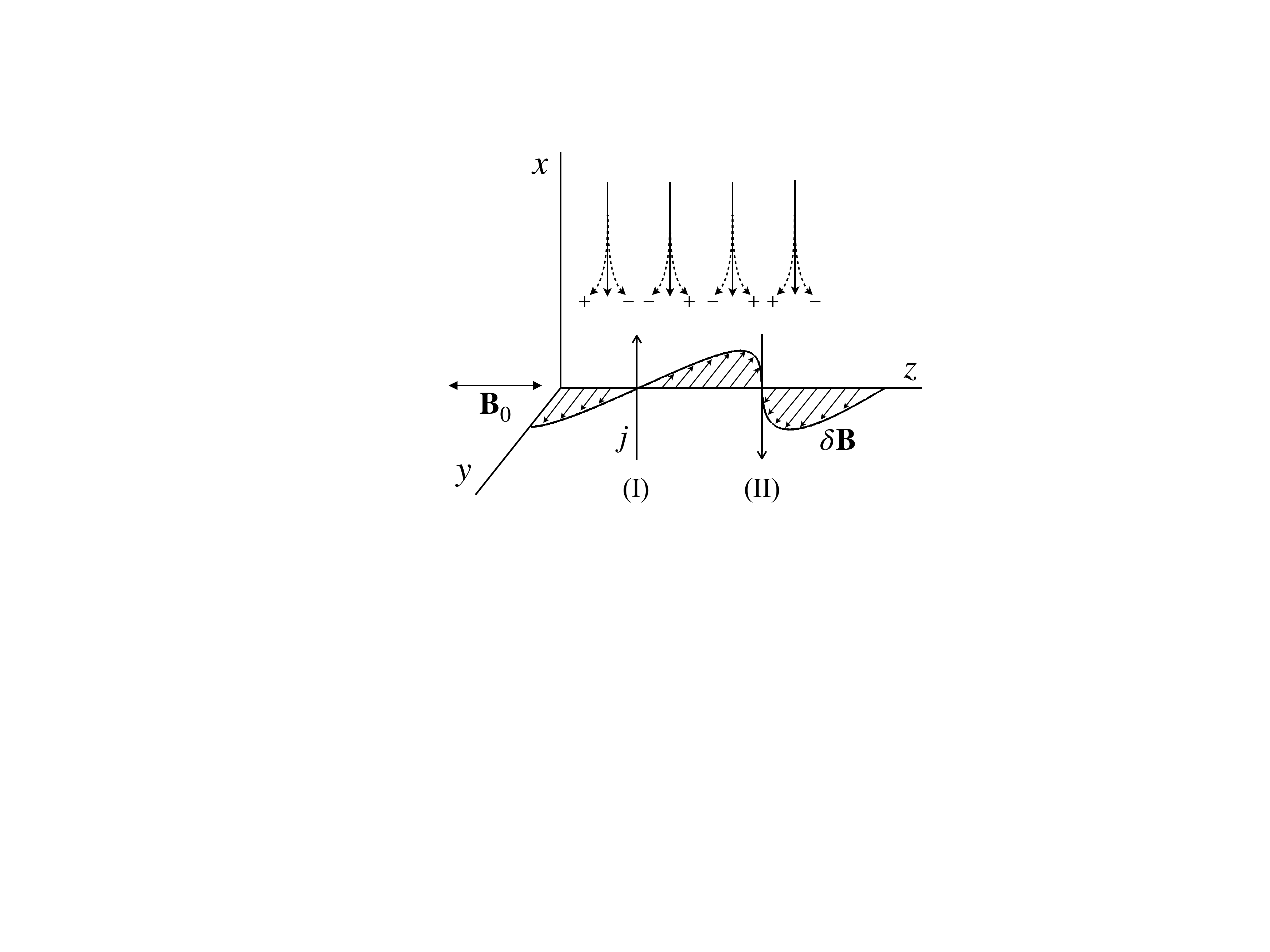}\\
	\caption{An illustration of the Weibel instability driven by the mDM beam in a perpendicular shock. The shock front faces toward the $\unitvec{x}$ direction. An initial magnetic fluctuation, $\delta {\mathbf B}$, is assumed to be perpendicular to ${\mathbf B}_{0}$ and lies in the $y-z$ plane. The incoming $\chi^\pm$s are deflected by $\delta {\mathbf B}$ and induce current sheets pointing in opposite directions in layers I and II. The magnetic perturbations are then amplified via Amp\`ere's Law.}
	\label{fig: diagram of Weibel-like}
\end{figure}

In the case of a perpendicular shock, we will show that an excess of kinetic energy from the mDM beam in the direction perpendicular to $\mathbf{B}_0$ can drive a Weibel instability~\cite{Weibel:1959hm} and excite EM waves. The Weibel instability is a transverse EM instability driven by a plasma with anisotropic velocity distribution. The instability can occur even in the absence of external magnetic field. The EM waves driven by the Weibel instability is \emph{nonresonant} with the particles. A simple physical picture of the Weibel instability is given in Ref.~\cite{Fried:1959gy} where the author treats a two-temperature Maxwellian gas as two counterstreaming cold plasmas. The same mechanism also applies to a single beam. In the following, we briefly describe the physical picture of the Weibel instability driven by the mDM beam, along the lines of the description in Refs~\cite{Fried:1959gy, Medvedev:1999tu}. 

We consider a neutral mDM beam consisting of equal numbers of $\chi^+$s and $\chi^-$s and with  beam velocity ${\mathbf V}_0 = -(3v_\sh/4)\unitvec{x}$ in the downstream plasma frame. An illustration is shown in Fig.~\ref{fig: diagram of Weibel-like}. Now consider an EM perturbation with $\mathbf{k} \parallel \mathbf{B}_0$ and $\delta{\mathbf B} \perp {\mathbf V}_0$. The Lorentz force $q_{\chi^\pm} {\mathbf V}_0\cross {\delta{\mathbf B}}$ deflects the mDM trajectories as shown by the dashed curves in Fig.~\ref{fig: diagram of Weibel-like}. At layer I, the $\chi^-$s congregate and $\chi^+$s disperse away, resulting an induced current sheet pointing in the $\unitvec{x}$ direction. The layer II has an opposite result, with the current sheet pointing in the $-\unitvec{x}$ direction. The magnetic fluctuations are then amplified via Amp\`ere's Law. Note that the fluctuations arising from this mechanism are aperiodic, i.e., $\omega_r \simeq 0$. 

We determine the growth rate via the linear stability analysis. The dispersion relation in the perpendicular shock is obtained by the same method as the previous subsection, with the exception that ${\mathbf V}_0 \perp {\mathbf B}_0$. The result is expressed as
\beq
\baln
    0 = D^\pm = c^2 k^2 - &\omega^2 - \sum_{j = i^+, e^-} \omega_{\plasma j}^2 \left(\frac{\omega}{k v_{\thermal,j}}\right) Z\left(\xi_j\right) \\
    &- \sum_{s=\chi^+,\chi^-} \omega_{\plasma s}^2\bigg[\left(\frac{\omega}{kv_{\thermal,\chi}}\right)Z\left(\xi_s\right) + \left(\frac{V_0}{v_{\thermal,\chi}}\right)^2\left(1 + \xi_s Z\left(\xi_s\right)\right)\bigg],
    \label{eq: Weibel-like dispersion relation}
\ealn    
\eeq
where $\xi_j = \left(\omega \pm \Omega_j \right)/kv_{\thermal,j}$ and $\xi_s = \left(\omega \pm \Omega_s\right)/ kv_{\thermal,\chi}$. The factor $V_0/v_{\thermal,\chi}$ quantifies the deviation of a plasma away from an  isotropic gas. For $V_0 = 0$, there is no free energy to drive the instability and Eq.~(\ref{eq: Weibel-like dispersion relation}) supports only stable solutions with $\gamma \leq 0$. The ions and electrons are again strongly magnetized ($|\omega|\ll |\Omega_i|, |\Omega_e|$ and $|\xi_i|,|\xi_e|\gg 1$) under the condition of small mDM Larmor frequency, $|\Omega_\chi/\Omega_i| < 10^{-3}$, and the low frequency, long wavelength EM waves that we are interested in. Thus, the thermal ions and electrons do not resonate with the EM waves and the cyclotron damping is exponentially suppressed. As a result, the sum of the ion and electron terms in Eq.~(\ref{eq: Weibel-like dispersion relation}) is reduced to $-\omega^2c^2/v_{\rm A}^2$ following the same reasoning discussed in Sec.~\ref{Sec: firehose}.

\begin{figure}[t!]
\centering
    \includegraphics[width=0.48\textwidth]{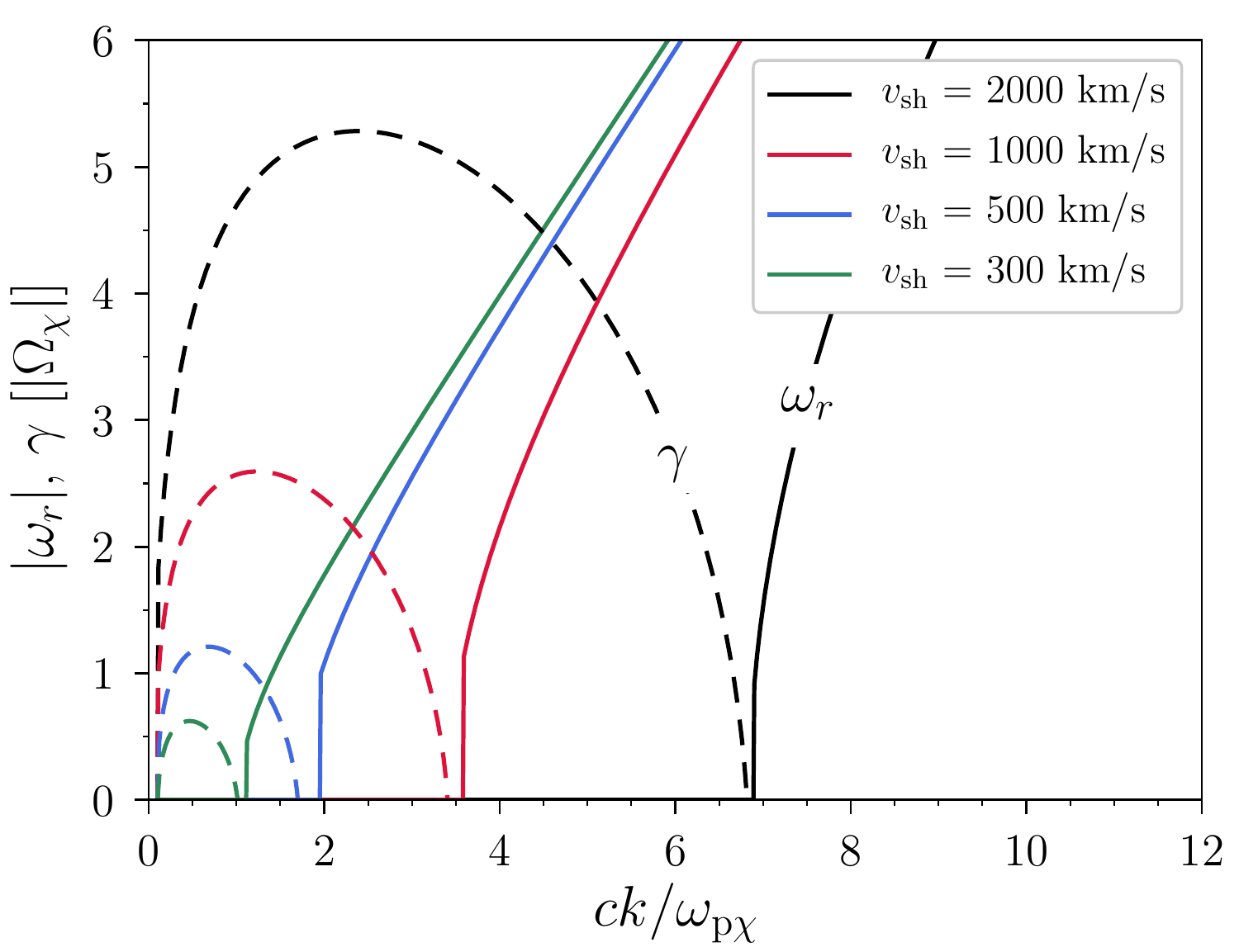}
    \includegraphics[width=0.48\textwidth]{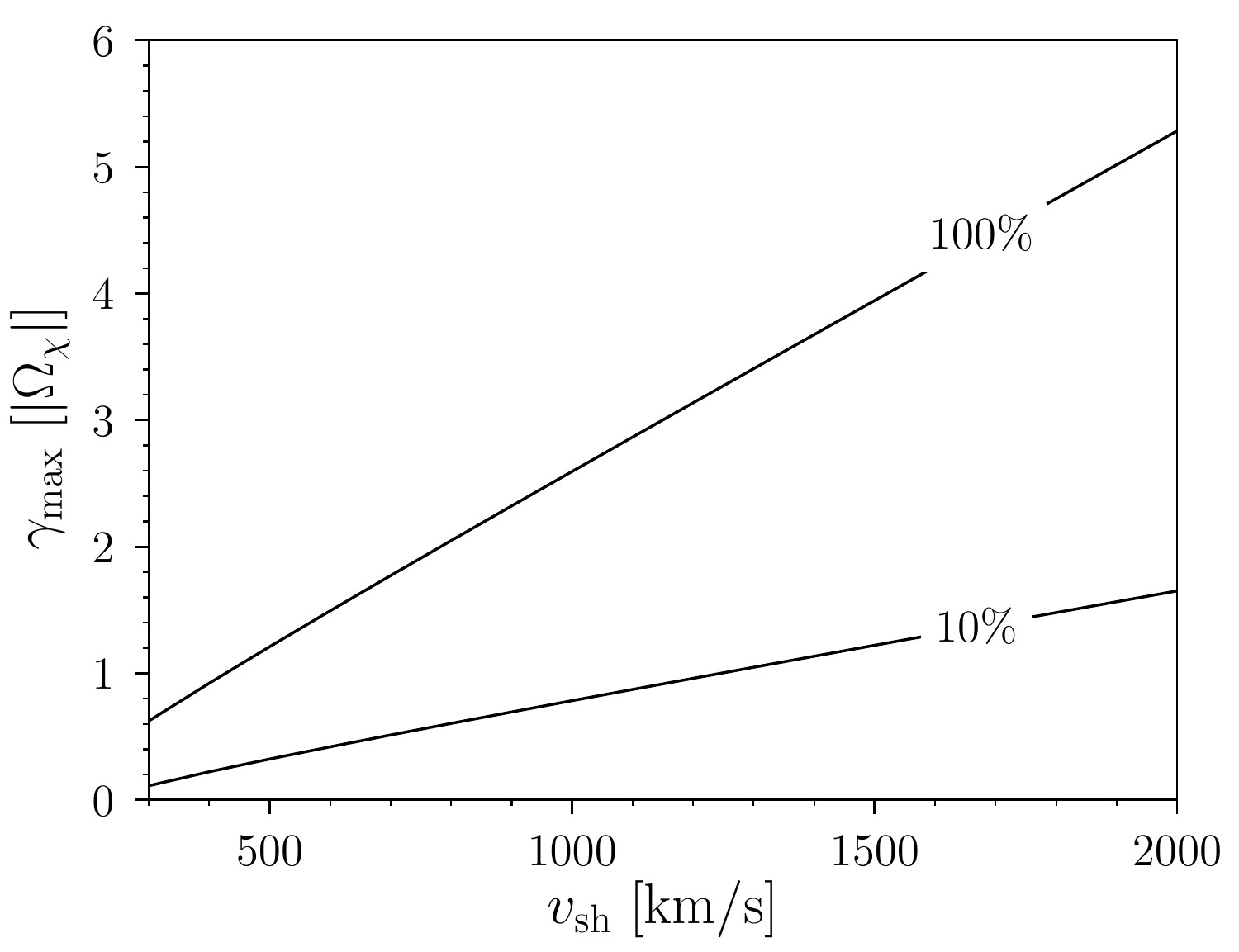}
    \caption{Weibel instability in a perpendicular shock. (left) Plot of the real oscillation frequency (solid lines) and the growth rate (dashed lines) of the right/left-hand EM wave as a function of wave vector, assuming $\rho_{\rm DM} = \rho_\chi = {\rm 1~GeV/cm^3}$. We note that one feature of Weibel instability is an aperiodic wave, i.e., $\omega_r \simeq 0$, which explains the vanishingly small values of $\omega_r$ for the unstable $k$ modes. (right) The maximum growth rate as a function of shock speed. The numbers on the curves denote $\rho_\chi/\rho_{\rm DM}$, assuming $\rho_{\rm DM} = {\rm 1~GeV/cm^3}$.}
    \label{fig: weibel_like_omega}
\end{figure}

The numerical solution of Eq.~(\ref{eq: Weibel-like dispersion relation}) is shown in Fig.~\ref{fig: weibel_like_omega} with the assumption $B_0 = 10~{\mu{\rm G}}$ and $n_{{\rm post},i} = 1~{\rm cm^{-3}}$. Both EM helicities have the same form of the dispersion relation. In the left panel of Fig.~\ref{fig: weibel_like_omega}, we show $|\omega_r|$ and $\gamma$ as a function of $ck/\omega_{\plasma \chi}$ for various $v_\sh$, assuming $\rho_{\rm DM} = \rho_{\chi} = {\rm 1~GeV/cm^3}$. Among the unstable $k$ modes, we find $\omega_r \simeq 0$. This indicates that the excited EM waves (left- and right-handed) are aperiodic and they barely propagate in space. The superposition of both helicities with the same phase then gives linearly polarized EM waves with $\omega_r \simeq 0$, which agrees with the features of the EM waves driven by the Weibel instability. We note that all curves of $\gamma$ eventually drop to zero or negative values as $k\to 0$ due to the fact that the free energy driving the instability is contained in the term $\left(V_0^2/v_{\thermal,\chi}^2\right)[1 + \xi_s Z\left(\xi_s\right)]$, which vanishes as $k \to 0$ ($\xi_s \to \infty$). In the right panel, we show the maximum growth rates as a function of $v_\sh$ with various values of $\rho_{\chi}/\rho_{\rm DM}$ assuming $\rho_{\rm DM} = {\rm 1~GeV/cm^3}$. The maximum growth rate $\gamma_{\rm max}$ is an increasing function of $v_\sh$ and $\rho_{\chi}/\rho_{\rm DM}$, and it is approximately proportional to $V_0/v_{\thermal,\chi}$. For the optimal scenario where $\rho_{\chi}/\rho_{\rm DM} = 100\%$, we find $\gamma \gtrsim |\Omega_{\chi}|$ for all shock speeds in the Sedov-Taylor phase ($v_\sh \gtrsim 200~\kms$).

We emphasize that while $\omega_r/k \rightarrow 0$ among the unstable $k$ modes shown in the left panel of Fig.~\ref{fig: weibel_like_omega}, the result does not suggest the kinetic waves in the plasma have zero phase speed. For the unstable EM fluctuations excited by the instability, the linear stability analysis performed here only contains the information about $\gamma$ and $\omega_r$ at a given $k$ mode as they are created. It does not provide information regarding the evolution and propagation of the fluctuations during the late stage of the instability. On the other hand, the kinetic plasma waves are the stable collective oscillations from tiny perturbations on an equilibrium plasma. Therefore, the phase speed of the growing fluctuations from instabilities should be distinguished from the phase speed of the kinetic waves propagating in the equilibrium plasma.

\subsection{Condition for sweeping up mDM}

We have shown that a mDM beam can drive the firehose and Weibel electromagnetic plasma instabilities, assuming a simplified model for the downstream plasma and magnetic field in a SNR. If the maximum growth rate is $\gamma_{\rm max}$, then the instability will be saturated in a timescale $\approx \gamma_{\rm max}^{-1}$. Consequently, for a shock of speed $v_\sh$, the saturation length in the downstream plasma is $\approx 3v_\sh \gamma_{\rm max}^{-1}/4$, where the mDM beam velocity is $V_0 = 3v_\sh/4$. That is, it takes about a distance scale of $3v_\sh \gamma_{\rm max}^{-1}/4$ to saturate the instability and make the mDM velocity distribution isotropic in the downstream frame. Since the majority of shocked ISM gas resides in the outermost shell with thickness $\approx 0.1 R_\sh$~\cite{draine2010physics}, we require that $3v_\sh \gamma_{\rm max}^{-1}/4 < 0.1 R_\sh$ in order for the mDM to be swept up and isotropized in an expanding SNR. 

The colored bands in Fig.~\ref{fig: mDM_parameter_space} show the resulting condition on the mDM charge and mass. We take the downstream ion density as $n_{{\rm post},i} = 1~{\rm cm^{-3}}$ and the ambient mDM mass density as $\rho_\chi = 1~{\rm GeV/cm^3}$. Above the orange (green) band, the mDM charge fraction $|q_\chi/e|$ is sufficiently large for it to drive a plasma instability and be swept up by the supernova in a parallel (perpendicular) shock. The upper and lower boundaries of each band are for $v_\sh = 1500~\kms$ and $300~\kms$, respectively, where we have used the appropriate value of the remnant radius $R_\sh$ for that shock speed (see Fig.~\ref{fig: remnant history}).

\section{Implications \label{sec:implications}}

In this section, we turn to some consequences of the mDM interaction in the SNR. Having seen that mDM can be swept up for sufficiently large charge fractions, we next discuss the eventual fate of the mDM velocity distribution once the shocked ISM gas undergoes adiabatic decompression and the supernova shock dies out. We comment on the possibility and challenges of generating a Fermi-accelerated mDM component in Sec.~\ref{subsec:dark cosmic rays}. Finally, the total mass swept up by the SNR can increase if mDM is accounted for, and we discuss how this could impact the Sedov-Taylor phase.

\subsection{Adiabatic decompression and the mDM velocity distribution}\label{sec: Adiabatic decompression}

When the mDM-driven plasma instabilities occur, a significant fraction of the ordered kinetic energy of the mDM beam is transferred to the disordered kinetic energy of the mDM particles moving isotropically. The mDM is ``shocked'' and heated up. For supernova shocks at the early and mid stages of the Sedov-Taylor phase, the swept-up mDM has bulk velocity (in the Milky Way frame) and velocity dispersion that are significantly larger than $v_{\thermal,\chi}\approx 220~\kms$, which would apparently have a large impact on terrestrial dark matter detection. However, it is not easy for these shocked mDM particles to stream out directly from the remnant for the following reason. After the instabilities have saturated, the mDM particles continue to scatter on the EM waves and are trapped in the supernova remnant. Consequently, we expect that the trapped mDM particles lose energy through adiabatic decompression as the shocked ISM gas returns to the ambient ISM state. At the same time, the bulk of the shocked mDM would slow down with the remnant and eventually come to rest in the Milky Way frame after the supernova shock has died out.

We first consider the effect of adiabatic decompression on the mDM velocity dispersion in the downstream frame.  As a simplified model\footnote{The exact mDM velocity distribution as well as the kinetic energy distribution between the shocked mDM particles and the mDM-driven waves can be obtained from the quasilinear theory or hybrid simulation.}, here we assume that (1) all the kinetic energy of the mDM beam measured in the downstream frame is transferred to the heat of  random mDM motion, (2) the process is instantaneous, i.e., the growth time is neglected, and (3) the shocked mDM particles follow a Maxwell-Boltzmann distribution after the instabilities have saturated. Then for a group of mDM particles shocked and swept up by the supernova shock wave with speed $v_\sh$, the resulting velocity dispersion is given as $v_{\rm dis,\chi}^\prime \left(v_\sh \right) = \sqrt{v_{\thermal,\chi}^2 + 3v_\sh^2/8}$. (Note that groups of mDM particles entering the shock at different shock speeds will have different velocity dispersion after being swept up. Since the different groups of mDM do not thermalize via Coulomb interactions for the parameters we study in this work, we use \emph{velocity dispersion} here to mean the dispersion for a given group of particles.)

Because the shocked ISM has a higher gas pressure than the ambient pressure, its volume expands. At the same time, the trapped mDM particles lose kinetic energy through adiabatic decompression. If the volume of the shocked ISM gas expands by a factor $\Lambda^{-3}$, then each mDM particles' speed is decreased by a factor $\Lambda$. That is, for a group of mDM particles swept up by the shock with speed $v_\sh$, the final velocity dispersion after decompression is given as $v_{\rm dis,\chi}^{\prime\prime}\left(v_\sh \right) = \Lambda v_{\rm dis,\chi}^{\prime}\left(v_\sh \right)$. Here we follow Ref.~\cite{Blandford:1980ja} to specify the decompression factor $\Lambda$ in two limits.

\emph{Density argument:} decompression stops once the shocked ISM gas is returned from its downstream density, $\rho_2$, to the ambient ISM density, $\rho_{\rm 1}$. According to the RH condition, $\rho_2 = 4 \rho_{\rm 1}$ for a high shock number. Thus, the decompression factor is
\beq
    \Lambda_d = \left(\frac{\rho_2}{\rho_{\rm 1}}\right)^{-1/3} = 4^{-1/3} \approx 0.63.
\eeq

\emph{Pressure argument:} decompression stops once the shocked ISM gas is returned from its downstream pressure, $P_2$, to the ambient ISM pressure, $P_{\rm 1}$. With the assumption that the shocked ISM gas is an ideal fluid and expands adiabatically, the decompression factor is
\beq
    \Lambda_p = \left(\frac{P_2}{P_{\rm 1}}\right)^{-1/5} =  \left(\frac{5M^2 -1}{4}\right)^{-1/5},
\eeq
where $M \equiv v_\sh / c_{\rm s}$ denotes the Mach number and $c_{\rm s}$ is the ambient ISM sound speed. For ISM temperature $T_{\rm 1} = 10^4~{\rm K}$ and adiabatic index $\gamma_{\rm a} = 5/3$ for a monatomic gas, the ISM sound speed is $c_{\rm s} = \sqrt{\gamma_{\rm a} T_{\rm 1}/m_i} \approx 12~\kms$.

\begin{figure}[t!]
\centering
    \includegraphics[width=0.49\textwidth]{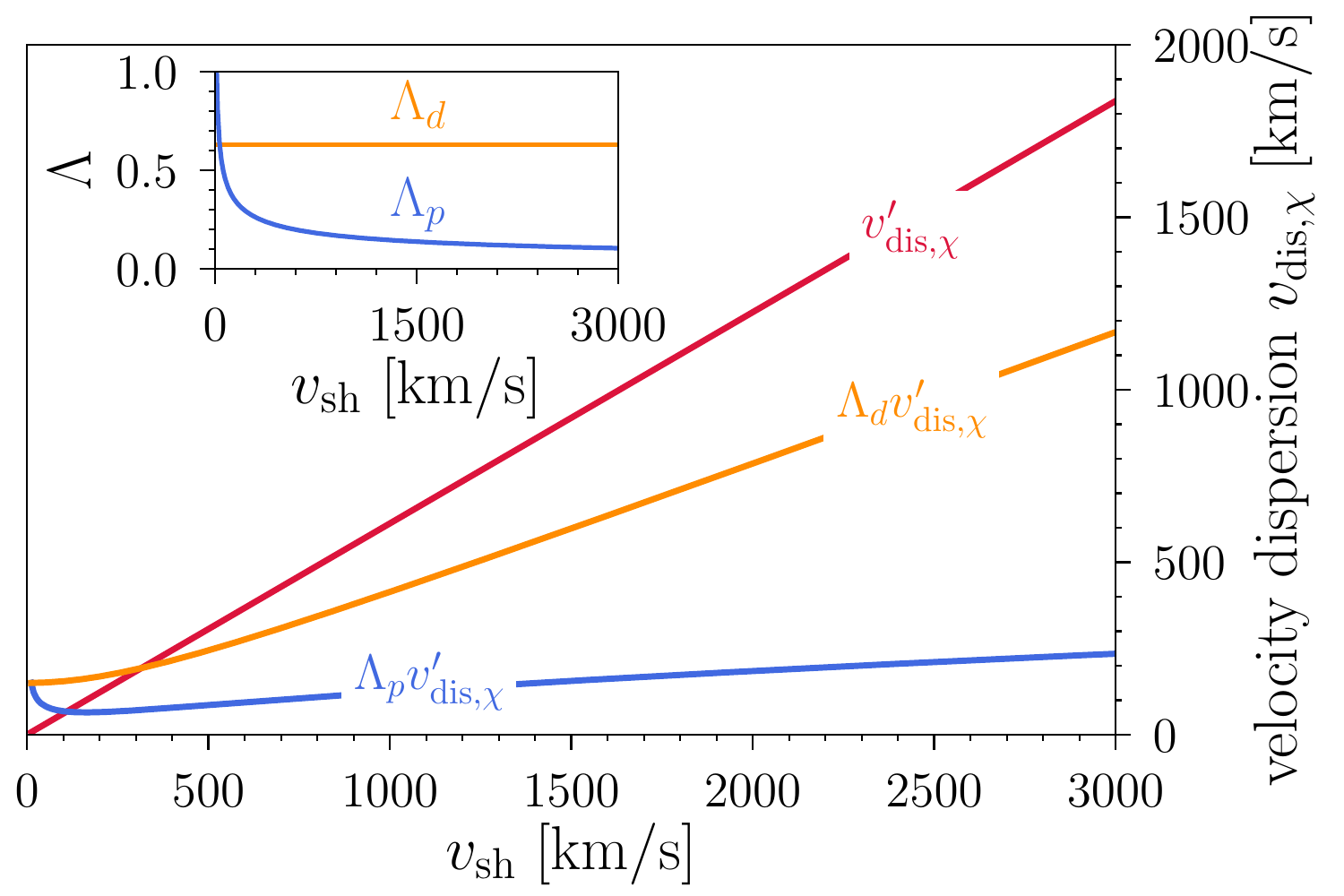}
    \caption{The effect of adiabatic decompression on the mDM velocity dispersion, in the frame of the downstream fluid. (main plot) The red dashed line is the velocity dispersion, $v_{\rm dis, \chi}^\prime$, for a group of mDM particles which have isotropized in a shock with speed $v_\sh$. Note that $v_{\chi}^{\prime}$ does not include any adiabatic decompression effect. The orange and blue lines are the mDM velocity dispersion after decompression, $v_{\rm dis, \chi}^{\prime\prime}\left(v_\sh\right) = \Lambda v_{\rm dis, \chi}^{\prime}\left(v_\sh\right)$, with $\Lambda$ given by a density ($\Lambda_d$) or  pressure ($\Lambda_p$) argument, 
    respectively. The actual velocity dispersion after decompression should lie between the orange and blue lines. (inset) Decompression factors $\Lambda_d$ and $\Lambda_p$. }
    \label{fig: adiabatic decompression}
\end{figure}

In Fig.~\ref{fig: adiabatic decompression}, we show the effect of adiabatic decompression on the shocked mDM particles. The orange and blue lines in the subplot denote $\Lambda_d$ and $\Lambda_p$, respectively. The decompression effect from the pressure argument is significantly stronger than the effect from the density argument. This is because the downstream pressure ($\sim \rho_{\rm 1} v_\sh^2$) is easily larger than the ambient ISM pressure ($\sim \rho_{\rm 1} c_{\rm s}^2$) by few orders of magnitude for high Mach shocks whereas the downstream density is always approximately 4 times the upstream density. The difference between the two arguments leads to a big contrast to the mDM velocity dispersion. In the main plot, the red dashed line represents the mDM velocity dispersion, $v_{\rm dis,\chi}^\prime$, for the group of mDM particles entering the shock with the speed $v_\sh$. The orange and blue lines represent the final velocity dispersion, $v_{\rm dis,\chi}^{\prime\prime}$, of the same group of mDM particles that have undergone $\Lambda_d$ and $\Lambda_p$, respectively. Note that $v_{\rm dis,\chi}^\prime$ always drops $\approx 40\%$ with the density argument whereas it can drop as much as $\gtrsim 80\%$ with the pressure argument for shocks in the Sedov-Taylor phase. A more realistic $v_{\rm dis,\chi}^{\prime\prime}$ within a galactic supernova remnant should lie between the orange and blue lines.

The mDM particles do not just free-stream out from the supernova remnant after adiabatic decompression has stopped. In the case of a quasiparallel shock, the mDM particles continue to scatter on the magnetic irregularities driven by the mDM beam or cosmic-ray protons. In the case of a quasiperpendicular shock, the mDM particles are confined in the downstream or near the shock front since the ordered upstream magnetic fields prohibit mDM streaming more than one Larmor radius. (Note that the latter case is independent of whether or not the trapping comes from mDM scattering with mDM-driven plasma waves.) In both scenarios, the shocked mDM particles remain trapped in the supernova remnant and the bulk velocity would slow down with the remnant. As a result, the velocity distribution of the shocked mDM in the Milky Way frame, $dN/dv_\chi$, is given as 
\beq
    \frac{1}{n_\chi} \frac{dN}{dv_\chi} = \int_{\rm 300~km/s}^{\rm 10^4~km/s} 4\pi R_\sh^2 \frac{dR_\sh}{dv_\sh} F_{\rm shell}\left(v_\chi, v_\sh\right) dv_\sh,
    \label{eq: number density function}
\eeq
where we have normalized by the ambient mDM number density, $n_\chi$, we have used $4\pi R_\sh^2 {dR_\sh}/{dv_\sh} = 2.94 E_{\rm SN} \rho_{\rm ISM}^{-1} v_\sh^{-3}$, and
\beq
    F_{\rm shell}\left(v_\chi, v_\sh\right) = \frac{4}{\sqrt{\pi}} \frac{v_\chi^2}{ ({v_{\rm dis,\chi}^{\prime\prime}}) ^3} e^{-v_\chi^2/({v_{\rm dis,\chi}^{\prime\prime}})^2}.
\eeq
Here as a conservative estimate in Eq.~(\ref{eq: number density function}), we only consider the supernova shock speed from $10^4~\kms$ to $300~\kms$, which is the region where mDM can drive plasma instabilities and be shocked in the case of $100\%$ dark matter content as mDM. (For $v_\sh \lesssim 300~\kms$, mDM beam does not drive EM plasma instabilities, as we have shown in Sec.~\ref{sec: EM instability}.)

\begin{figure}[t]
\centering
    \includegraphics[width=0.49\textwidth]{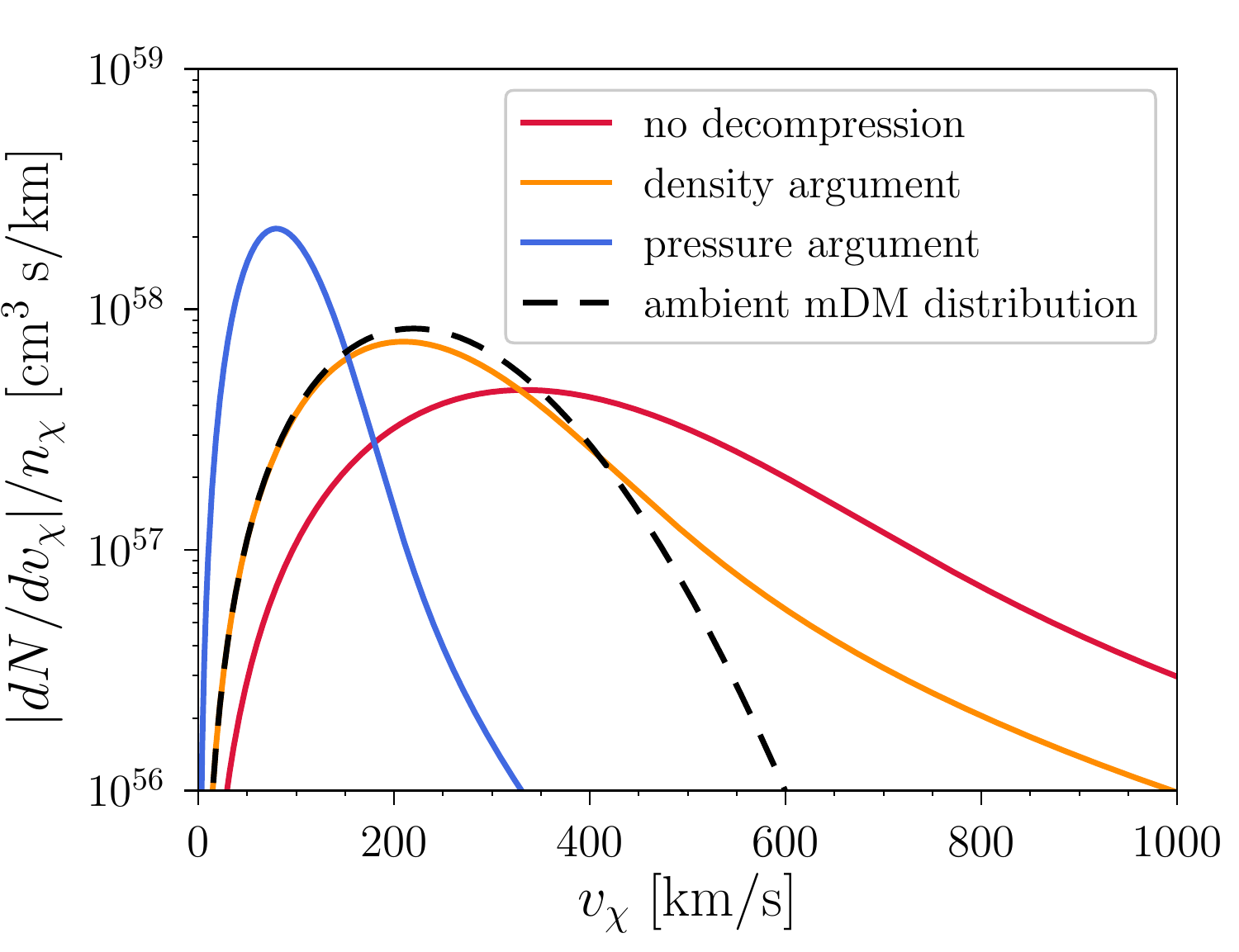}
    \caption{The velocity distribution of the shocked mDM in each supernova event. We only consider mDM particles entering the supernova shock of speed between $10^4~\kms$ and $300~\kms$. The black dashed curve is the ambient (unshocked) mDM distribution. While decompression according to the density argument only moderately cools down the shocked mDM, decompression from the pressure argument can make the majority of the shocked mDM colder than the ambient mDM.}
    \label{fig: distribution_decompression}
\end{figure}

In Fig.~\ref{fig: distribution_decompression}, we show the velocity distribution in the cases with and without adiabatic decompression. The red curve does not include any decompression which means $v_{\rm dis,\chi}^{\prime\prime} = v_{\rm dis,\chi}^{\prime}$. The orange and blue curves include adiabatic decompression with the density argument and the pressure argument, respectively. The black dashed curve is for mDM that has not interacted with any SNR. The tail of the red curve comes from the swept-up mDM in the early phase of the Sedov-Taylor phase where the shock speed is high. It does not result from the DSA mechanism. The decompression from density argument only decreases the tail of high-speed particles moderately since the shocked ISM density is only 4 times the ambient ISM density and the bulk of the fluid is not decompressed much. On the other hand, the decompression from the pressure argument can cool down the shocked mDM significantly, even colder than the ambient mDM. We expect that the shocked mDM should lie between these limits. These arguments suggest that, after being shocked and cooled by decompression, the bulk of the mDM velocity distribution ends up qualitatively similar to the initial ambient mDM distribution.

\subsection{Challenges for generating dark cosmic rays \label{subsec:dark cosmic rays} }

We have shown that mDM can be shocked and swept up by the supernovae. A natural question one might ask is whether mDM also undergoes the DSA process and is Fermi-accelerated to  relativistic speeds in the same way cosmic-ray protons are accelerated. Having relativistic mDM particles in the Milky Way would have a significant impact on terrestrial dark matter detection~\cite{Hu:2016xas, Dunsky:2018mqs}. However, there is no consensus on the flux of Fermi-accelerated mDM in the Milky Way. Reference~\cite{Dunsky:2018mqs} considers (1) the possibility of getting pre-accelerated mDM from one shock and injecting them into the DSA process in another shock, as well as (2) the possibility that mDM particles with the virial velocity larger than the shock speed can undergo Fermi acceleration in one shock if they are never thermalized with the ISM gas. On the other hand, Ref.~\cite{Hu:2016xas} assumes that the mDM can be injected into the DSA process in just one shock if the mDM particles have the same rigidity as some of the Fermi-accelerated protons in the same shock. Because the two works assume very different acceleration processes, the resulting Fermi-accelerated fluxes are incompatible. Nevertheless, there are still several critical issues that must be answered. In this section, we bring up the challenges of getting Fermi-accelerated mDM taking into consideration plasma waves and adiabatic decompression.

The standard acceleration theory utilizes the diffusion-convection equation in a quasiparallel shock. It requires that (1) the accelerated particle speed be much higher than the shock speed and (2) the phase space distribution of the accelerated particles in the local fluid frame is isotropic. Then in the steady state, the diffusion-convection equation gives the isotropic part of the phase space distribution in the downstream, $f_+$, in terms of the isotropic part of the phase space distribution in the far upstream, $f_{-}$, in the following form~\cite{Drury:1983hp, Blandford:1987bi, Malkov:1995wd}
\beq
    f_{+}\left(p\right) = s p^{-s} \int^p dp^\prime f_{-}\left(p^\prime\right) {p}^{\prime s-1} + C/p^s,
    \label{eq: standard acceleration theory equation}
\eeq
where $s = 3r/(r-1)$ decides the slope of the Fermi-accelerated spectrum and $r \approx 4$ denotes the compression ratio of the flow given by the Rankine-Hugoniot conditions. The first term on the right-hand side (rhs) of Eq.~(\ref{eq: standard acceleration theory equation}) provides the spectrum of the Fermi-accelerated particles in the downstream when the preexisting energetic particles in the far upstream are advected through the shock front and undergo the DSA process. The second term represents the spectrum of the particles that are directly injected from the downstream particles into the DSA process. The constant $C$ is determined by matching the spectrum of the injected particles at the supra-thermal regime\footnote{Supra-thermal regime is the transition zone where the distribution of the downstream thermal particles is joined to the power-law distribution of the Fermi-accelerated particles.} with the power-law spectrum. 

In Ref.~\cite{Dunsky:2018mqs}, the authors use the first term on the right-hand side of Eq.~(\ref{eq: standard acceleration theory equation}) to obtain a Fermi-accelerated mDM spectrum. The proposed physical origins are the following: (1) the mDM with $m_\chi/q_\chi^2 < 3 \times 10^6 \: {\rm GeV}/e^2$ are thermalized with ISM. To Fermi-accelerate them, two shocks are required. The moving magnetic fields near the shock front of the first supernova shock reflect mDM particles and accelerate them to the shock speed. Among the reflected mDM, all the particles moving faster than $200~\kms$ can then undergo DSA in a second shock with $v_\sh = 200~\kms$. This process is referred as \emph{two-stage injection} in Ref.~\cite{Dunsky:2018mqs}. (2) The mDM with $m_\chi/q_\chi^2 > 3 \times 10^6 \: {\rm GeV}/e^2$ are not thermalized with ISM and therefore have the velocity of the Milky Way virial velocity (taken as $220~\kms$). On encountering a shock with $v_\sh = 200~\kms$, all mDM particles are Fermi-accelerated because they move faster than the shock. Here we revisit both mechanisms, taking into consideration plasma waves.

We first review the idea of particle reflection at the shock front from the standard cosmic ray acceleration theory. It was known that a fraction of upstream protons may be reflected by the potential barrier at the shock front due to the shock reformation~\cite{Kucharek:1991el, Guo:2013hmy, Caprioli:2014dwa}. However, heavy ions with mass-to-charge ratio larger than protons are not reflected by the shock barrier; instead they penetrate to the downstream~\cite{Caprioli:2017oun}. Since the mDM we study here has mass-to-charge ratio significantly larger than the ratio of protons, we also expect mDM impinge to the downstream directly, which represents the first difficulty for the two-stage mechanism. The mDM particles flow into the downstream fluid and are swept up by the SNR. After that, they likely still remain trapped in the downstream fluid due to the wave-particle scattering. They will undergo adiabatic decompression and slow down along with the expanding SNR. The number density of the energetic mDM (say with speeds significantly higher than $200~\kms$ in the Milky Way frame) from the first shock will be reduced after taking into consideration these effects, as shown in Fig.~\ref{fig: distribution_decompression}.

This leads to the second challenge of getting DSA in the two-stage injection process. For the second supernova shock with $v_\sh = 200~\kms$, the first term on the rhs of Eq.~(\ref{eq: standard acceleration theory equation}) operates exclusively to preaccelerated particles with speeds \emph{at least} several thousand $\kms$. Such high speeds can be achieved only if mDM enters the first supernova shock at the free-expansion phase or the very early stage of the Sedov-Taylor phase so that the catastrophic energy loss from the adiabatic decompression is overcome. Consequently, the total number of these high-speed mDM from the first supernova shock which might undergo standard acceleration in the second shock would certainly be several orders of magnitude lower than the total number of the mDM particles covered within $ R_\sh\left(200~\kms\right)$ of the first supernova shock. Taking all these issues into consideration, we find it challenging to get Fermi-accelerated mDM from multiple supernova shocks. 

The direct injection of the downstream mDM particles to the DSA process is another possibility to get dark cosmic rays, as proposed in Refs.~\cite{Dunsky:2018mqs, Hu:2016xas}. However, studying this mechanism from first principles is a difficult problem, and generally simulations are required to determine the constant $C$ in Eq.~(\ref{eq: standard acceleration theory equation}). (Ref.~\cite{Dunsky:2018mqs} assumes an $\mathcal{O}\left(1\right)$ injection fraction.) There are multiple stages to obtaining the injection rate. Here we only qualitatively discuss three challenges/issues, while much more work is required before obtaining a self-consistent picture of the mDM injection rate. The very first thing to figure out is the exact mDM velocity distribution in the downstream, since it controls the number of downstream mDM particles that can return to upstream after they have passed the shock for their first time. In this work, we merely use linear stability analysis to understand the timescale for mDM to be swept up by the SNR. The mDM velocity distribution can be obtained by utilizing quasilinear theory, which describes the phase space evolution and the beam relaxation back to a marginally stable state~\cite{diamond2010modern}. 

The second challenge is understanding the dynamics of the mDM particles that return back to the upstream from the downstream. In the case of proton cosmic ray acceleration, it is believed that the protons returning to the upstream from the downstream (or reflected at the shock front back upstream) scatter on self-generated Alfv\'{e}n waves propagating along the ambient magnetic fields in the (quasi)parallel shock. Those protons are then isotropized in the upstream frame and can return downstream again. While the same strategy can be applied to mDM, it is not clear how the injection process is altered due to the ultraslow instabilities of mDM particles compared to that of cosmic-ray protons. 

The third issue is that the Alfv\'{e}n waves excited in the upstream will eventually cross the shock front and trap a fraction of mDM particles trying to escape to the upstream. This effect reduces the odds that particles can gain energy from multiple crossings at the shock front and be injected to DSA process~\cite{Kato:2014ewa, Marcowith:2016vzl}. Different velocity distribution functions will lead to different probabilities that particles can evade the trapping. It is therefore important to understand how the downstream mDM with a non-Maxwellian distribution responds to the particle-trapping from the upstream Alfv\'{e}n waves. Addressing all these questions is needed to determine the normalization and slope of any Fermi-accelerated mDM component.

\subsection{Sedov-Taylor phase of the supernova remnant}

If the mDM particles are swept up by the supernova remnant, the total mass density of the swept-up fluid, $\rho_0$, is increased. Then we expect the shock wave to propagate slower and travel a shorter distance than the case in which only the ISM is swept up. In principle, we can determine this slowing-down effect by measuring $R_\sh$, $\dot{R}_\sh$ and $\ddot{R}_\sh$ of the younger Type Ia supernovae, which are at the early stage of the Sedov-Taylor phase and where the dynamics of expansion is predominantly governed by the swept-up mass. We can infer the effect of deceleration from the swept-up mDM once the $E/\rho_0$ is obtained, where $E$ is the initial kinetic energy of the ejecta.

A simple estimate shows that it is not easy to make an observable prediction, however. The typical ISM mass density is $\rho_{\rm ISM} \sim 1~{\rm GeV/cm^3}$. In the optimal case where mDM constitutes all of the DM, the mDM mass density at few kpc away from the Galactic Center (GC) is also $\rho_{\chi}\sim  1~{\rm GeV/cm^3}$. Thus, the total swept-up mass $\rho_0 = \rho_{\rm ISM} + \rho_{\chi}$ is at most a few times larger. On the other hand, the kinetic energy of the ejecta from a Type Ia supernova is approximately $1-2\times10^{51}~{\rm erg}$ -- which is also uncertain by a factor a few. With these uncertainties from the mass density and the ejecta kinetic energy, it is challenging to quantify the effect of deceleration from the swept-up mDM or make any constraints on the local mDM density. Lastly, we note that while the volume within a few hundred pc from GC has $\rho_{\rm DM} > 10~{\rm GeV/cm^3}$ and is an ideal site to test the deceleration effect, there are no observed supernovae in that region within the past 1000 years. The lack of events is due to the fact that the molecular clouds in the central molecular zone (inner $\sim200~{\rm pc}$) is subject to disruptive shear which suppresses the star formation rate~\cite{Kauffmann:2017dw}.

\section{Conclusion \label{sec:conclusion}}

We have shown that it is possible for mDM to be swept up by supernova remnants and to become isotropized in the expanding fluid through plasma instabilities. A few of the most representative plasma waves and instabilities are considered: ion-acoustic, Langmuir, and lower-hybrid wave instabilities of electrostatic waves, and beam-firehose and Weibel instabilities of electromagnetic waves. While there is a tremendous amount of free energy from the relative motion between the ambient mDM and the supernova remnants, we find that the electrostatic waves are never excited due to the fact that mDM with $|q_\chi|/m_\chi \ll e/m_p$ is not able to drive the growth fast enough to overcome ion Landau damping. On the other hand, electromagnetic waves can be excited by mDM because the wave frequency is much smaller than the ion cyclotron frequency and thus the ion cyclotron damping is avoided. 

We find that for $|q_\chi/m_\chi| \gtrsim 10^{-13} e / {\rm MeV}$, mDM can be isotropized in the expanding supernova remnants at the Sedov-Taylor phase by driving electromagnetic beam-firehose and Weibel instabilities in parallel and perpendicular shocks, respectively, provided that mDM constitutes all the DM content in the Milky Way. If mDM makes up only a fraction of the total DM, then mDM might not able to drive electromagnetic plasma waves at the late stage of the Sedov-Taylor phase since the mDM beam velocity threshold for the instability increases as the density of mDM decreases. 

Finally, we emphasize that the plasma waves prevent the majority of the shocked mDM particles from free-streaming out the supernova remnant via wave-particle interactions. Moreover, the mDM particles undergo significant energy loss from  adiabatic decompression as the remnant expands. Both effects play important roles in the final velocity distribution of the shocked mDM in the Milky Way, and our analysis suggests that the bulk of the shocked mDM ends up with a velocity distribution qualitatively similar to the unshocked distribution. Further study of mDM wave-particle interactions is needed to determine to what extent mDM undergoes Fermi-acceleration and its subsequent Galactic dynamics.

\begin{acknowledgments}
We thank Tsang Keung Chan, Hope How-Huan Chen, Daniel Dubin, David Dunsky, Dušan Kereš, Robert Lasenby, Diego Redigolo and Kurt Thompson for valuable discussions.
We especially appreciate the guidance on plasma physics and cosmic rays from Patrick Diamond and Mikhail Malkov. J.T.L. is supported by an Government Scholarship to Study Abroad from Taiwan government. T.L. is supported by an Alfred P. Sloan Research Fellowship and Department of Energy (DOE) grant no. \textrm{DE-SC0019195}.
\end{acknowledgments}

\appendix
\section{REVIEW OF UNMAGNETIZED LINEAR RESPONSE FUNCTION}\label{appendix: Landau dielectric function}
In this Appendix, we follow closely the derivations in Refs.~\cite{krall1973principles, galeev1983basic} to briefly review the dispersion relation for an unmagnetized plasma with no magnetic fields present. The derivation of the linear response for a magnetized plasma is similar, and can be found in many plasma physics books, e.g., Refs~\cite{krall1973principles, galeev1983basic, gary2005theory, ichimaru2018basic}.

We start with the collisionless Vlasov-Maxwell equations
\beq
\baln
    \left(\frac{\pd}{\pd t} + \mathbf{v}\cdot\nabla + \frac{q_j}{m_j} \mathbf{E}\cdot \frac{\pd}{\pd \mathbf{v}}  \right) f_j\left(\mathbf{x}, \mathbf{v}, t\right) = 0, \qquad &\text{Vlasov equation}, \\
    \nabla\cdot \mathbf{E} = \sum_j 4 \pi q_j \int f_j\left(\mathbf{x}, \mathbf{v}, t\right) d^3 \mathbf{v}, \qquad &\text{Gauss Law},
\ealn
\eeq
where $\mathbf{E} = \mathbf{E}\left(\mathbf{x}, t\right)$ is the electric field, and $q_j$, and $m_j$ are the charge and  mass of species $j$. 
In the present analysis, we consider longitudinal waves with small amplitude propagating through a system near equilibrium.
We consider a spatially homogeneous plasma with no external electric and magnetic fields, i.e., $\mathbf{E}_0 = \mathbf{B}_0 = 0$. Define a normalized particle distribution function of the species $j$ as 
\beq
    F_j\left(\mathbf{x}, \mathbf{v}, t\right) \equiv \frac{f_j\left(\mathbf{x}, \mathbf{v}, t\right)}{n_{0j}} ,
\eeq
where $f_j\left(\mathbf{x}, \mathbf{v}, t\right)$ is the density in ($\mathbf{x}, \mathbf{v}$) phase space and $n_{0j}$ is the ambient number density. Both $\mathbf{E}\left(\mathbf{x}, t\right)$ and $F_j\left(\mathbf{x}, \mathbf{v}, t\right)$ can be written as the sum of their equilibrium values plus small perturbation terms:
\beq
\baln
    \mathbf{E}\left(\mathbf{x}, t\right) &\rightarrow \delta \mathbf{E}\left(\mathbf{x}, t\right), \\
    F_j\left(\mathbf{x}, \mathbf{v}, t\right) &\rightarrow F_{0j}\left(\mathbf{v}\right) + \delta F_j\left(\mathbf{x}, \mathbf{v}, t\right).
\ealn    
\eeq
The analysis is greatly simplified if we apply the Fourier-Laplace transform on $\delta \mathbf{E}$ and $\delta F_j$,
\beq
\baln
    \delta \mathbf{E}\left(\mathbf{x}, t\right) &= \int d^3 \mathbf{k} \: e^{i \mathbf{k}\cdot \mathbf{x}} \int_{\omega_0 - i \infty}^{\omega_0 + i \infty} \frac{d\omega}{2\pi i} e^{ \omega t} \delta \mathbf{\tilde{E}}\left(\mathbf{k}, \omega \right),\\
    \delta F_j\left(\mathbf{x}, \mathbf{v}, t\right) &= \int d^3 \mathbf{k} \: e^{i \mathbf{k}\cdot \mathbf{x}} \int_{\omega_0 - i \infty}^{\omega_0 + i \infty} \frac{d\omega}{2\pi i} e^{ \omega t} \delta \tilde{ F}_j\left(\mathbf{k}, \mathbf{v}, \omega \right),
\ealn
\eeq
where $\omega_0 = {\rm Re}\left(\omega \right)$ is chosen large enough so that the reverse integrals $\int_0^\infty \delta \mathbf{\tilde{E}}\left(\mathbf{k}, \omega \right) \exp(-\omega t) \: dt$ and $\int_0^\infty \delta {\tilde{F_j}}\left(\mathbf{k}, \mathbf{v}, \omega \right) \exp(-\omega t) \: dt$ converge. Rewriting the linearized Vlasov-Maxwell equations in  ($\mathbf{k},\omega$) space, then
\begin{align}
    -i \left(\omega - \mathbf{k}\cdot\mathbf{v}\right) \delta \tilde{F}_j \left(\mathbf{k}, \mathbf{v}, \omega \right) = \frac{q_j}{m_j} \delta \mathbf{\tilde{E}} \cdot \frac{\pd F_{0j}\left(\mathbf{v}\right)}{\pd \mathbf{v}},
    \label{eq: Vlasov, kw} \\
    i \mathbf{k}\cdot \delta\mathbf{\tilde{E}} \left(\mathbf{k}, \omega \right) = \sum_j 4\pi n_{0j} q_j \int \delta \tilde{F}_j \left(\mathbf{k}, \mathbf{v}, \omega\right) d^3 \mathbf{v}.
    \label{eq: Gauss, kw}
\end{align}
It is straightforward to show from Eqs.~(\ref{eq: Vlasov, kw}) and (\ref{eq: Gauss, kw}) that
\beq
     i \mathbf{k}\cdot \delta\mathbf{\tilde{E}} \left(\mathbf{k}, \omega \right) \bigg[ 1 + \sum_j \frac{\omega_{\plasma j}^2}{k^2} \int \frac{\mathbf{k} \cdot \pd F_{0j}\left(\mathbf{v}\right)/\pd \mathbf{v}}{\omega - \mathbf{k}\cdot\mathbf{v}} \: d^3 \mathbf{v} \bigg] = 0,
     \label{eq: nontrivial solution}
\eeq
where $\omega_{\plasma j}^2 = 4\pi n_{0j} q_j^2/m_j$. Equation~(\ref{eq: nontrivial solution}) applies for the longitudinal component of $\delta \tilde{\bf E}$.
The requirement for a nontrivial solution of Eq.~(\ref{eq: nontrivial solution}) is 
\beq
    D\left(\mathbf{k}, \omega\right) \equiv 1 + \sum_j \frac{\omega_{\plasma j}^2}{k^2} \int \frac{\mathbf{k} \cdot \pd F_{0j}\left(\mathbf{v}\right)/\pd \mathbf{v}}{\omega - \mathbf{k}\cdot\mathbf{v}} \: d^3 \mathbf{v} = 0,
    \label{eq: dispersion relation, unmagnetized, general}
\eeq
Equation~(\ref{eq: dispersion relation, unmagnetized, general}) is the definition of the unmagnetized plasma dielectric function, $D\left(\mathbf{k}, \omega\right)$, and the associated dispersion relation.

\subsection{Plasma dispersion function}\label{subsec: plasma dispersion function}
In many plasma environments, each species is modeled as a gas in thermal equilibrium. It is therefore useful to further simplify the dispersion relation in Eq.~(\ref{eq: dispersion relation, unmagnetized, general}). A drifting Maxwellian gas of species $j$ has the distribution function given by
\beq
    F_{0j}\left(\mathbf{v}\right) = \pi^{-3/2} \: v_{\thermal, j}^{-3} \: \exp\left(-\frac{\left(\mathbf{v} - \mathbf{V}_{0j}\right)^2}{v_{\thermal, j}^2}\right),
    \label{eq: Maxwell distribution}
\eeq
where $\mathbf{V}_{0j}$ and $v_{\thermal,j} = \sqrt{2T_j/m_j}$ are the drift velocity and the thermal velocity of the species $j$, respectively. Using Eq.~(\ref{eq: Maxwell distribution}) in Eq.~(\ref{eq: dispersion relation, unmagnetized, general}), the dispersion relation of the unmagnetized plasma is expressed as
\beq
    0 = D\left(\mathbf{k}, \omega\right) = 1 + \sum_j \frac{2\omega_{\plasma j}^2}{k^2 v_{\thermal,j}^2} \big[1 + \xi_j Z\left(\xi_j\right)\big],
    \label{eq: dispersion relatio for Maxwellian gas}
\eeq
where $\xi_j = \left(\omega - \mathbf{k} \cdot \mathbf{V}_{0j}\right)/k v_{\thermal, j}$ and 
\beq
    Z\left(\xi_j\right) \equiv \frac{1}{\sqrt{\pi}} \int_{-\infty}^{\infty} \frac{e^{-x^2}}{x - \xi_j} \: dx.
    \label{eq:Zdef}
\eeq
Here we examine some important properties of the plasma dispersion function widely used in the literature of plasma physics. The asymptotic expansion of $Z\left(\xi_j\right)$ for small and large values of $|\xi_j|$ are given as~\cite{fried1961plasma}
\beq
    Z\left(\xi_j\right) = i \sqrt{\pi} \exp\left(- \xi_j^2\right) - 2 \xi_j + \frac{4}{3} \xi_j^3 - \frac{8}{15}\xi_j^5 + \cdots, \qquad \text{for}~ |\xi_j|< 1,
    \label{eq: Z, small xi}
\eeq
and
\beq
    Z\left(\xi_j\right) = i \sqrt{\pi} \exp\left(- \xi_j^2\right) - \frac{1}{\xi_j} - \frac{1}{2\xi_j^3} - \frac{3}{4\xi_j^5} + \cdots \qquad \text{for}~ |\xi_j|> 1.
    \label{eq: Z, large xi}
\eeq
Also, differentiating $Z\left(\xi_j\right)$ with respect to $\xi_j$ and applying integration by parts, we obtain
\beq
    Z^\prime\left(\xi_j\right) = \frac{-2}{\sqrt{\pi}} \int_{-\infty}^{\infty} \frac{xe^{-x^2}}{x - \xi_j} \: dx = -2 \big[ 1 + \xi_j Z\left(\xi_j\right)\big] .
    \label{eq: Z prime identity}
\eeq

As an example to show the application of Eqs.~(\ref{eq: dispersion relatio for Maxwellian gas}),~(\ref{eq: Z, small xi}) and~(\ref{eq: Z, large xi}), consider Langmuir waves propagating through a neutral electron-proton plasma at equilibrium with zero drift velocities. Landau damping is avoided if we require that $\omega_r/k \gg v_{\thermal, e} \gg v_{\thermal, i}$, i.e., $|\xi_e| \gg 1$ and $|\xi_i| \gg 1$. The dispersion relation in Eq.~(\ref{eq: dispersion relatio for Maxwellian gas}) becomes
\beq
\baln
    0 = D\left(\mathbf{k}, \omega \right) &= 1 - \frac{\omega_{\plasma e}^2}{\omega_r^2} - 3k^2 \lambda_{\Debye e}^2 \frac{\omega_{\plasma e}^4}{\omega_r^4} - \frac{\omega_{\plasma i}^2}{\omega_r^2} + \cdots\\
    &\approx 1 - \frac{\omega_{\plasma e}^2}{\omega_r^2} - 3k^2 \lambda_{\Debye e}^2 \frac{\omega_{\plasma e}^4}{\omega_r^4} + \mathcal{O}\left(\omega_{\plasma e}^6\right),
\ealn    
\eeq
where $\lambda_{\Debye e} \equiv \sqrt{T_e/4\pi n_e e^2}$ is the electron Debye screening length. The ion contribution is neglected because $\omega_{\plasma e}^2 \gg \omega_{\plasma i}^2$ in the neutral plasma. In the limit $k^2 \lambda_{\Debye e}^2 \ll 1$ where electron Landau damping is avoided, we obtain the dispersion relation for stable Langmuir waves,
\beq
    \omega_r = \omega_{\plasma e} \sqrt{1 + 3 k^2 \lambda_{\Debye e}^2}.
\eeq

Note that the Landau damping contribution comes from the term $i \sqrt{\pi} \exp\left(-\xi_j^2\right)$ in Eqs.~(\ref{eq: Z, small xi}) and~(\ref{eq: Z, large xi}). In the regime where $\xi_j \sim 1$, the exponential term is not negligible, and the waves suffer strong Landau damping from the species $j$. To get the damping rate, one can numerically solve Eq.~(\ref{eq: dispersion relatio for Maxwellian gas}) or use the technique shown in the following subsection.

\subsection{Weakly damped or growing waves}\label{subsection: Weakly damped or growing waves}
In this section, we provide some useful results on the growth rate for the weakly damped or growing waves. Expanding $D\left(\mathbf{k}, \omega_r + i\gamma \right)$ in the limit $|\gamma/\omega_r| \ll 1$, we obtain 
\beq
    0 = D\left(\mathbf{k}, \omega_r + i \gamma \right) = D_{\rm Re}\left(\mathbf{k}, \omega_r\right) + i D_{\rm Im}\left(\mathbf{k}, \omega_r\right) + i \gamma \frac{\pd D_{\rm Re}\left(\mathbf{k}, \omega_r\right)}{\pd \omega_r} ,
    \label{eq: Taylor-expansion on dispersion relation}
\eeq
where we have separated $D\left(\mathbf{k}, \omega_r + i \gamma \right)$ into real and imaginary parts and assumed higher order terms are small. The real part of Eq.~(\ref{eq: Taylor-expansion on dispersion relation}), $D_{\rm Re}\left(\mathbf{k}, \omega_r\right) = 0$, determines the real oscillation frequency. Setting the imaginary part to zero then gives the damping or growing rate,
\beq
    \gamma = \frac{-D_{\rm Im}\left(\mathbf{k}, \omega_r\right)}{\pd D_{\rm Re}\left(\mathbf{k}, \omega_r\right)/\pd \omega_r},
    \label{eq: growth rate equation}
\eeq
where $\gamma < 0$ and $\gamma > 0$ correspond to damping and growing waves, respectively.

On the other hand, the dielectric function $ D\left(\mathbf{k}, \omega_r + i \gamma \right)$ can be directly decomposed into its real and imaginary components by utilizing the Plemelj formula,
\beq
    \frac{1}{\omega_r - \mathbf{k} \cdot \mathbf{v} + i \gamma}\bigg\rvert_{|\gamma| \ll \omega_r} = \mathcal{P} \frac{1}{\omega_r - \mathbf{k} \cdot \mathbf{v}} - i \pi \delta\left(\omega_r - \mathbf{k} \cdot \mathbf{v}\right) ,
\eeq
with $\mathcal{P}$ denoting the Cauchy principle value. We obtain 
\begin{subequations}
\begin{align}
    D_{\rm Re}\left(\mathbf{k}, \omega_r\right) = 1 + \sum_j \frac{\omega_{\plasma j}^2}{k^2} \: \mathcal{P} \int \mathbf{k}\cdot \frac{\pd F_{0j} \left(\mathbf{v}\right)}{\pd \mathbf{v}} \frac{1}{\omega_r - \mathbf{k} \cdot \mathbf{v} } d^3 \mathbf{v},
    \label{eq: D_RE}\\
    D_{\rm Im}\left(\mathbf{k}, \omega_r\right) = -\pi \sum_j \frac{\omega_{\plasma j}^2}{k^2} \int   \mathbf{k}\cdot \frac{\pd F_{0j} \left(\mathbf{v}\right)}{\pd \mathbf{v}} \: \delta \left( \omega_r - \mathbf{k} \cdot \mathbf{v} \right) \: d^3 \mathbf{v}.
     \label{eq: D_IM}
\end{align}
\label{eq: D_REIM}
\end{subequations}

\bibliographystyle{apsrev4-1}
\bibliography{mDM}

\end{document}